\documentclass[11pt,reqno]{article}

\usepackage{jheppub}

\makeatletter
\def\@fpheader{Preprint typeset in JHEP style}
\makeatother
\usepackage{mathtools}

\usepackage{url}
\usepackage{hyperref}

\definecolor{darkBlue}{rgb}{0,0,0.5}
\hypersetup{colorlinks=true,linktocpage=true,linkcolor=darkBlue,citecolor=darkBlue,urlcolor=darkBlue}

\usepackage{slashed}

\newtheorem{theorem}{Theorem}[section]
\newtheorem{lemma}[theorem]{Lemma}

\newcommand\be{\begin{equation}}
\newcommand\ee{\end{equation}}
\def\ba#1\ea{\begin{align}#1\end{align}}
\def\nba#1\nea{\begin{align*}#1\end{align*}}
\newcommand\bi{\begin{itemize}}
\newcommand\ei{\end{itemize}}
\newcommand\ben{\begin{enumerate}}
\newcommand\een{\end{enumerate}}

\newcommand{\R}{\ensuremath{\mathbb{R}}}
\newcommand{\CC}{\ensuremath{\mathbb{C}}}
\newcommand{\Z}{\ensuremath{\mathbb{Z}}}

\newcommand{\Mcal}{\mathcal{M}}

\newcommand{\foh}{\frac{1}{2}}

\newcommand{\mattwo}[4]
{
\begin{pmatrix}
                        #1  & #2   \\
                        #3 &  #4
                          \end{pmatrix} }

\newcommand{\sym}{\mathsf{SYM}}
\newcommand{\msym}{\mathsf{MSYM}}

\newcommand{\Dbar}{\overline{\mathcal{D}}}
\newcommand{\Ncal}{\mathcal{N}}
\newcommand{\Lcal}{\mathcal{L}}
\newcommand{\thetabar}{\overline{\theta}}

\DeclareMathOperator{\Tr}{Tr}

\DeclareMathOperator{\Sym}{Sym}
\DeclareMathOperator{\Hom}{Hom}

\def\JKRes{\mathop{\mbox{JK-Res}}\limits}
\def\Res{\mathop{\mbox{Res}}\limits}

\newcommand{\Qcal}{\mathcal{Q}}

\numberwithin{equation}{section}

\begin{document}

\title{Quantum Vacua of 2d Maximally Supersymmetric Yang-Mills Theory}

\author{Murat Kolo\u{g}lu}
\affiliation{Walter Burke Institute for Theoretical Physics, California Institute of Technology, Pasadena, CA 91125, USA}


\abstract{
We analyze the classical and quantum vacua of 2d $\mathcal{N}=(8,8)$ supersymmetric Yang-Mills theory with $SU(N)$ and $U(N)$ gauge group, describing the worldvolume interactions of $N$ parallel D1-branes with flat transverse directions $\mathbb{R}^8$. 
We claim that the IR limit of the $SU(N)$ theory in the superselection sector labeled $M \pmod{N}$ --- identified with the internal dynamics of $(M,N)$-string bound states of the Type IIB string theory --- is described by the symmetric orbifold $\mathcal{N}=(8,8)$ sigma model into $(\mathbb{R}^8)^{D-1}/\mathbb{S}_D$ when $D=\gcd(M,N)>1$, and by a single massive vacuum when $D=1$, generalizing the conjectures of E. Witten and others. 
The full worldvolume theory of the D1-branes is the $U(N)$ theory with an additional $U(1)$ 2-form gauge field $B$ coming from the string theory Kalb-Ramond field. This $U(N)+B$ theory has generalized field configurations, labeled by the $\mathbb{Z}$-valued generalized electric flux and an independent $\mathbb{Z}_N$-valued 't Hooft flux. 
We argue that in the quantum mechanical theory, the $(M,N)$-string sector with $M$ units of electric flux has a $\mathbb{Z}_N$-valued discrete $\theta$ angle specified by $M \pmod{N}$ dual to the 't Hooft flux. Adding the brane center-of-mass degrees of freedom to the $SU(N)$ theory, we claim that the IR limit of the $U(N) + B$ theory in the sector with $M$ bound F-strings is described by the $\mathcal{N}=(8,8)$ sigma model into ${\rm Sym}^{D} ( \mathbb{R}^8)$.
We provide strong evidence for these claims by computing an $\mathcal{N}=(8,8)$ analog of the elliptic genus of the UV gauge theories and of their conjectured IR limit sigma models, and showing they agree. 
Agreement is established by noting that the elliptic genera are modular-invariant Abelian (multi-periodic and meromorphic) functions, which turns out to be very restrictive. 
}

\maketitle




\section{Introduction and summary}

Supersymmetric Yang-Mills theories ($\sym$) have been of central interest in string theory, especially since the advent of D-branes. 
In Type II string theories, the worldvolume interactions of BPS D$p$-branes at low energies are described by maximally supersymmetric Yang-Mills theories in $(p+1)$-dimensions ($\msym_{p+1}$). 
These theories have $16$ supersymmetries, inherited from the target-space supersymmetries left unbroken by the half-BPS D-branes. For a stack of $N$ D-branes, the gauge group of the $\msym$ is $U(N)$. The gauge field arises from the open strings that stretch between pairs of branes, which carry $U(N)$ Chan-Paton factors when the branes are coincident. The gauge theory is enhanced by the higher-form gauge fields and fluxes present in the string theory target place, which generalize the topological sectors of the theory.
Properties of these gauge theories are intimately related to the interactions of D-branes. For example, topological sectors of the gauge theory are interpreted as the bound states of the branes with other objects in the string theory, including other  D-branes of various dimensions and the fundamental string \cite{Witten:1995im}. In fact, an entire non-perturbative formulation of M-theory was conjectured to arise from the $N\rightarrow \infty$ limit of the $\Ncal=16$ quantum mechanics $\msym_1$ describing the interactions of D0-branes \cite{Banks:1996vh}.

In this article, we will focus on the 
two-dimensional (2d) $\msym$ 
theories with gauge group $U(N)$ or $SU(N)$. In two dimensions, the weakly coupled gauge theory defined by the $\sym$ Lagrangian is inherently the ultraviolet (UV) description, and such theories are asymptotically free. In the infrared (IR), the theory becomes strongly coupled. It is a difficult and interesting question to understand the infrared dynamics of $\msym_2$. Both of the closely related theories with $U(N)$ and $SU(N)$ gauge group have been extensively analyzed, and much has been conjectured about their infrared description and quantum vacua \cite{Witten:1995im,Dijkgraaf:1997vv,Motl:1997th,Seiberg:1997ax,F.Antonuccio:1998aa}. 
For example, in \cite{Dijkgraaf:1997vv,Motl:1997th}, 
$U(N)$ $\msym_2$ theory 
was developed into matrix string theory, describing matrix theory compactified on a circle. It was proposed that the $N\rightarrow \infty $ limit of this theory should provide a non-perturbative formulation of Type IIA string theory. Using M-theory and string duality considerations, the authors of \cite{Dijkgraaf:1997vv} related the IR limit of $\msym_2$ with gauge group $U(N)$ to the supersymmetric sigma model into the symmetric orbifold $\Sym^N \R^8$, identified as the sector of second quantized free Type IIA strings with light-cone momentum $p_+=N$. 
However, exact computations or quantitative evidence have been elusive --- a situation we seek to remedy. 

The Lagrangian of $\msym_2$ can be obtained by dimensional reduction from 10d $\Ncal=1$ $\sym$, and for $U(N)$ or $SU(N)$ gauge group it is given by \cite{Dijkgraaf:1997vv,F.Antonuccio:1998aa}
\be
\Lcal = \Tr\left( -\frac{1}{4} F_{\mu\nu}^2  -\frac{1}{2} (D_\mu X^i)^2 + i\chi^T \slashed{D} \chi +\frac{g^2}{4}[X^i,X^j]^2 -\sqrt{2} g \chi_L^T \gamma_i[X^i,\chi_R] \right). \label{eq:msymlag}
\ee The bosons $X^i$, the left-moving fermions $\chi_L^{\dot\alpha}$, and the right moving fermions $\chi_R^{\alpha}$ are in the $\mathbf{8}_v$, $\mathbf{8}_c$, and $\mathbf{8}_s$ representations, respectively, of the $Spin(8)$ R-symmetry. The fields are also in the adjoint representation of the gauge group, so they are valued in $\mathfrak{u}(N)$ ($\mathfrak{su}(N)$) and can be realized as $N\times N$ (traceless) Hermitian matrices for gauge group $U(N)$ ($SU(N)$). The theory has $\Ncal=(8,8)$ supersymmetry generated by the transformations with 16 fermionic parameters $(\epsilon_L^{\alpha},\epsilon_R^{\dot\alpha})$. We take the worldsheet directions to be $\mu = 0, 9$. The dimensional reduction of the Lagrangian and the supersymmetry transformations are reproduced in Appendix \ref{apx:dim red and susy}.

The $\msym_2$ theory was observed to have classical vacua determined by the zeroes of the bosonic potential $V(X) =\frac{g^2}{4}[X^i,X^j]^2$, which are commuting matrices $X^i$, modulo the Weyl group $\mathbb{S}_N$ permuting the eigenvalues \cite{Witten:1995im,Dijkgraaf:1997vv}. For the $U(N)$ theory on the worldsheet $\R_t \times S^1$, all the zero-energy configurations of the gauge field correspond to flat connections on the trivial $U(N)$-principal bundle, so in the quantum $U(N)$ theory, the gauge field contributes a single trivial zero-energy state to the vacuum wavefunction, as elaborated in \cite{Dijkgraaf:1998zd}. Therefore, it seems natural to conjecture that in the infrared limit, as $g\rightarrow \infty$, the theory flows to the supersymmetric sigma model into $\Sym^N(\R^8)$, parametrized by the $N$ eigenvalues of the $X^i$ and fermionic partners \cite{Dijkgraaf:1997vv}. Similar arguments could be made for the $SU(N)$ theory, by removing the contributions for the free diagonal $U(1)$ factor of the $U(N)$ theory, leading to the supersymmetric sigma model into $(\R^8)^{N-1}/\mathbb{S}_N$ as the conjectural IR limit.

However, this is not all of the vacua and therefore not the end of the story. 
In his analysis of bound states of fundamental strings and D-branes in Type II string theories, Witten \cite{Witten:1995im} argued that the existence of $(M,N)$-string bound states in Type IIB string theory requires the existence of various supersymmetric vacua for the $SU(N)$ $\msym_2$. For the worldvolume theory of $N$ D1-branes, the sector with $M$ bound fundamental strings corresponds to a ``charge at infinity'' in the form of a Wilson loop in the $M$th tensor power of the fundamental representation of $SU(N)$ \cite{Witten:1995im}. Therefore, the $(M,N)$-string is naturally a superselection sector in the 2d quantum theory, and the vacuum in that sector is identified as the discrete $\theta$ vacuum \cite{Witten:1978ka} (of the related $SU(N)/\Z_N$ theory)  
with $\theta$ angle specified by $M \pmod{N}$ as \begin{equation} e^{i\theta} = e^{i\frac{2\pi M}{N}}. \end{equation} 
Specifically, Witten argued that the case when $M$ and $N$ are relatively prime should correspond to a single supersymmetric vacuum of the $SU(N)$ theory with a mass gap. This is because the center-of-mass motion of the branes decouples from the $U(N)$ worldvolume theory as a free $\Ncal =(8,8)$ $U(1)$ vector multiplet, corresponding to the determinant $U(1)$ in $U(N)$ (which decomposes as $U(N)= (U(1) \times SU(N))/\Z_N$). In the case with $M$ and $N$ relatively prime, the center-of-mass dynamics encoded in the decoupled $U(1)$ multiplet correspond to all of the massless physical degrees of freedom of the bound state in the string theory target space. 

In the more general case when $M$ and $N$ are not relatively prime, Witten reasoned that there is no argument to indicate the corresponding vacuum should be massive. In fact, the $(M,N)$-string should be able to split up into $D$ many $(M/D,N/D)$-string bound states without an energy barrier, where $D =\gcd(M,N)$, as the eigenvalues of the scalars corresponding to the relative positions of these $(M/D,N/D)$-strings can take arbitrary expectation values at no cost in energy. It is then natural to expect that the vacuum corresponding to the $(M,N)$-string with $D>1$ should have massless excitations corresponding to the massless degrees of freedom of the relative motion of the $(M/D,N/D)$-strings. The relative positions of these bound states is just the configuration space of $D$ indistinguishable strings in the transverse space $\R^8$, with the center-of-mass moduli excluded, which is described by the 2d symmetric orbifold sigma model into $(\mathbb{R}^8)^{D-1}/\mathbb{S}_D$.

We would like to analyze the classical and the quantum theory, and determine to what extent these predictions hold. The main feature of the $\msym_2$ theory which gives rise to some important subtleties is that all the local fields are in the adjoint representation of the gauge group $G$. In particular, if $G$ has a nontrivial center $Z(G)$, then there are no fields charged under it, so the $Z(G)$ charge cannot be screened, giving rise to superselection sectors labeled by the $Z(G)$ charge. For example, for $G=SU(N)$, $Z(G)=\Z_N$, and there are $N$ superselection sectors. Given a state in some sector, the emanation of a Wilson loop in some representation $R$ of $SU(N)$ with charge $N_R$ under $\Z_N$ will yield a state in another superselection sector, differing by $N_R$ units modulo $N$. Since there are no fields charged under the center, we can also define the $G/Z(G) = SU(N)/\Z_N$ theory that has the same Lagrangian. The $SU(N)/\Z_N$ theory has the $\theta$ angle parameter as additional discrete data, and for each of the $N$ choices of $\theta$, the spectrum is a restriction of the $SU(N)$ spectrum to one of the $N$ superselection sectors. Likewise, one can define the $SU(N)/\Z_K$ theory for $K|N$, which will have $N/K$ superselection sectors with the same $\Z_K\subset \Z_N$ charge for each of the $K$ choices of the 
$\theta$ angle.

Interestingly, when we consider the classical vacua of the $SU(N)/\Z_N$ theory, we recover a spectrum consistent with the spectrum of relative positions of the $(M/D,N/D)$-strings. This requires analyzing the topological sectors of the theory. 
Let's recall that the discrete $\theta$ vacua exist for the $SU(N)/\Z_N$ theory because 
this gauge group has non-trivial fundamental group $\pi_1(SU(N)/\Z_N) =\Z_N$. Consequently, there are ``instanton sectors'' of the 2d theory corresponding to the topologically distinct $SU(N)/\Z_N$-principal bundles, labeled by elements in $\pi_1(SU(N)/\Z_N)$ \cite{Witten:1978ka, Hooft:1979uj}. We denote the $\Z/N\Z$-valued instanton number by $k$. As usual, the effect of the $\theta$ angle in the path integral is to weigh the $k$-instanton sector by $e^{i\theta k}$ in the sum over the instanton sectors.
Naturally, the $\theta$ angle takes values in the Pontryagin dual of the $\pi_1$ of the gauge group, which is $\Z_N$ once again for $\pi_1(SU(N)/\Z_N)=\Z_N$. The theory at a given $\theta$ angle could be explicitly defined by including a surface operator constructed from the integral of a 2-form gauge field, as in \cite{Kapustin:2014gua}.
When one puts the $SU(N)/\Z_N$ theory on the two-torus $T^2$, the $SU(N)/\Z_N$-principal bundle $P_{N,k}$ over $T^2$ with instanton number $k$ admits flat connections, with moduli space $\mathcal{M}_{N,k}$, so there are classical zero-energy configurations of the gauge field in each instanton sector. As all of the fields are in one $\Ncal=(8,8)$ vector multiplet, the modes supersymmetric to the zero-energy modes of the gauge field are also classically zero-energy field configurations. The moduli space of flat connections $\mathcal{M}_{N,k}$ turns out to have complex dimension $d-1$, where $d=\gcd(k,N)$. Thus, 
one expects on general supersymmetry grounds to have a $8(d-1)$ real dimensional moduli space of vacua for the scalar fields, specifically $(d-1)$ real moduli for the eigenvalues of each of the scalars $X^i$. Indeed, when $d=1$, $\mathcal{M}_{N,k}$ is a point, and there is a single classical zero-energy field configuration with all the scalars set to zero. When $d>1$, the zero-energy scalar fields take the form
\begin{equation} X^i = I_{N/d} \otimes \begin{pmatrix} x_1^i & & \\ & \ddots & \\ & & x_d^i \end{pmatrix}, \qquad  \text{with } \Tr X^i = 0, \label{eq:classical vacua in instanton sector d}
\end{equation} 
in the strong coupling limit $g \rightarrow \infty$, and the eigenvalues parametrize $(\mathbb{R}^8)^{d-1}/\mathbb{S}_d$.  When $d=N$, we are in the trivial instanton sector with $k=0$, with the classical vacua described by $(\mathbb{R}^8)^{N-1}/\mathbb{S}_N$, in agreement with \cite{Witten:1995im,Dijkgraaf:1997vv}.

In the quantum theory, the wavefunction of a vacuum state spreads over all classical vacuum configurations, including the disconnected components. 
Although one expects that the quantum vacua should parallel the classical vacua in theories with high supersymmetry, one might be hesitant to reach this conclusion in our setting as it is \emph{a priori} unclear how the sum over classical disconnected configurations reproduces the vacua wavefunctions. 
Nonetheless, the $\theta$ angle isolates superselection sectors corresponding to $(M,N)$-strings, which have string theoretic descriptions strikingly in parallel with the classical vacua, supporting this conclusion. Here, a few relevant studies are crucial in guiding one's intuition. First of all, the $SO(8)$ R-symmetry anomalies vanish for $\msym_2$ \cite{Witten:1995im}, so there are no anomaly arguments that rule out the existence of the various massive and massless vacua, unlike in theories with less supersymmetry. Also, in \cite{Seiberg:1997ax}, it was argued that the IR description of $\msym_2$ could not be a non-trivial superconformal field theory with $\Ncal=(8,8)$ supersymmetry, as there is no extension of this $\Ncal=(8,8)$ supersymmetry algebra to a linear \emph{superconformal} algebra \cite{Nahm:1977tg}.\footnote{Non-linear $\Ncal=8$ superconformal algebras have been constructed, however they are quite exceptional and do not seem to be relevant to $\msym_2$. See \cite{Gunaydin:1995as} and the references therein for details.} This suggests that any scale invariant theories with massless excitations describing the IR fixed points should be free theories, or orbifolds thereof.
Lastly, in \cite{F.Antonuccio:1998aa}, $\msym_2$ was analyzed using discrete light-cone quantization (DLCQ). There, numerical results were obtained in finite resolution of light-cone momentum indicating the absence of normalizable massless states and supporting the existence of a vacuum with mass gap for the $SU(N)$ theory. 
By these considerations, the only possible choices for the IR limit of $\msym_2$ are massive vacua or orbifolds of free $\Ncal=(8,8)$ sigma models.
Given the favorable evidence, we conjecture that the quantum vacuum of the $SU(N)/\Z_N$ theory with $\theta = 2\pi M/N$ corresponding to the $(M,N)$-string should be described by the sigma model into $(\mathbb{R}^8)^{D-1}/\mathbb{S}_D$, and furthermore that the infrared fixed point of the theory with the given $\theta$ angle is this sigma model. We note that this description is invariant under the $SL(2,\Z)$ S-duality of the Type IIB string theory, which acts on the doublet $(M,N)$ but leaves $D$ invariant. Also, the vacua of the related $SU(N)$ theory in one of its $N$ superselection sectors is the vacuum of the $SU(N)/\Z_N$ theory with the corresponding $\theta$ parameter.

We provide strong evidence in favor of our claim by computing the $\Ncal=(8,8)$ analog of the elliptic genus --- or, index for short 
 --- of $\msym_2$ for $SU(N)$ and $SU(N)/\Z_N$ gauge group, for the latter also including the surface operator 
specifying the $\theta$-angle parameter. This index is a supersymmetric partition function on the Euclidean flat torus $T^2$ (with conformal class $\tau$), which counts states that are BPS with respect to a conjugate pair of right-moving supercharges. The choice of any such supercharge commutes with a $Spin(6)$ subgroup of the $Spin(8)$ R-symmetry, and we can refine the index with equivariant parameters $a_{1,2,3}=\exp {2\pi i \xi_{1,2,3}}$ coupling to the $Spin(6)$ subgroup. This refinement keeps track of more information about the spectrum, as well as regulating the otherwise divergent sum over the infinitely many states contributed by the non-compact bosonic zero-modes. This index also agrees with the equivariant elliptic genus of the theory when viewed as a $\Ncal=(0,2)$ supersymmetric theory --- from which perspective the $Spin(6)$ symmetry is just a flavor symmetry. Concretely, the index of an $SU(N)/\Z_K$ theory is defined as the following trace in the Ramond-Ramond (RR) Hilbert space $\mathcal{H}$ of the theory, which is a direct sum of $K$ RR Hilbert spaces on the circle, $\mathcal{H}_k$, quantized in the given instanton background $k$:
\begin{align} \mathcal{I}^\theta(\tau|\xi) 
&= \sum_{k}  e^{i\theta k} \Tr_{\mathcal{H}_{k}} (-1)^F a^f q^{H_L} \bar{q}^{H_R} . \end{align} Here, $q=e^{2\pi i \tau}$, and $H_L$ and $H_R$ are the left- and right-moving Hamiltonians.
We show that the index of the $SU(N)/\Z_N$ theory with the $\theta(M) = 2\pi M/N$ vacuum is \begin{equation} \mathcal{I}_{SU(N)/\Z_N}^{\theta(M)} ( \tau | \xi) = \frac{\mathcal{I}_D}{\mathcal{I}_1} (\tau| \xi), \label{eq:index matching} \end{equation} where $D = \gcd(M,N)$, and $\mathcal{I}_D$ is the index of the supersymmetric sigma model into $\Sym^D(\R^8)$. Of course, when $D=1$, $\mathcal{I}_{SU(N)/\Z_N}^\theta=1$, which is the index of a single massive supersymmetric vacuum. When $D>1$, $\mathcal{I}_D / \mathcal{I}_1$ is the index of the sigma model into $(\R^8)^{D-1}/\mathbb{S}_D$, since by factoring the diagonal copy of $\R^8$, we have $\Sym^D(\R^8) = \R^8 \times (\R^8)^{D-1}/\mathbb{S}_D$. 
The expressions for $\mathcal{I}_{SU(N)/\Z_N}^\theta ( \tau | \xi)$ and $\frac{\mathcal{I}_D}{\mathcal{I}_1} (\tau| \xi)$ are obtained through different methods, and it is non-trivial to show that they agree. Thankfully, both sets of functions enjoy multi-periodicity and $SL(2,\Z)$ modular invariance, and using these very restrictive properties we are able to establish \eqref{eq:index matching} for $N\le 7$. Since the index is an invariant of the theory under renormalization group (RG) flow, which is furthermore a ``strong'' invariant in the sense that it contains data about the spectrum of the theory, matching the index computed in the UV with the index of our candidate IR fixed point is a powerful indication that the two theories are indeed related by RG flow. 

From the $SU(N)/\Z_N$ index, we infer the index of the $SU(N)/\Z_K$ theory for any $K|N$,
\begin{equation}
\mathcal{I}_{SU(N)/\Z_K}^{\theta(M)} ( \tau | \xi) = \sum_{m\equiv M \pmod{K}} \frac{\mathcal{I}_{\gcd(m,N)}}{\mathcal{I}_1} (\tau| \xi),
\end{equation} where the sum is over the $N/K$ values of positive integers $m$ between $1$ and $N$ equivalent to $M$ modulo $K$. The terms being summed over are interpreted as the indices of the corresponding superselection sectors of the theory, and they are consistent with our earlier analysis of the superselection sectors.


Having understood the vacua $SU(N)$ $\msym_2$, we would like to analyze the $U(N)$ theory as well.  
Including the center of mass modes into our considerations of the $SU(N)$ theory, one can readily conjecture that the $U(N)$ theory also has vacua described by sigma models into $\Sym^D(\R^8)$ corresponding to the $(M,N)$-strings
, as expected from string theory. However, the correct analysis of the full N D1-brane worldvolume theory is somewhat more complicated, and requires some discussion. For a standard 2d $U(N)$ gauge theory with only adjoint fields, the $U(1)$ degrees of freedom decouple, and the index of the standard $U(N)$ $\msym_2$ can be readily inferred from the $SU(N)$ index as 
\begin{equation}
\mathcal{I}_{U(N)} (\tau| \xi) = \mathcal{I}_{U(1)}\mathcal{I}_{SU(N)} (\tau| \xi) = \sum_{m=1}^N \mathcal{I}_{\gcd(m,N)} (\tau| \xi).
\end{equation} But, this theory is not accurately taking into account the full structure of the $(M,N)$-string bound states. 
The true gauge theory describing the full worldvolume theory of the $N$ D1-branes is not a standard $U(N)$ gauge theory, but also has the Kalb-Ramond 2-form gauge field $B$ coming from the Neveu-Schwarz (NS) sector of the string theory. The $B$-field has an Abelian gauge symmetry generated by a 1-form gauge transformation, under which the trace mode of the $U(N)$-connection $A$ is also charged. Due to this additional 1-form gauge symmetry, the theory has generalized field content roughly described by $U(1)\times SU(N)/\Z_N$ gauge bundles, and the structures of the classical and quantum vacua are different. Indeed, we find that the $U(N)$ $\msym_2$ with the 2-form $B$-field has 
sectors corresponding to the $(M,N)$-strings as sought. The $M$th sector has a net $U(1)$ generalized electric flux of $M$ units, which is interpreted as the flux of the $M$ F-strings, as well as a $\theta$ angle $2\pi M/N$ in the $SU(N)/\Z_N$ sector. When $M=0$, the net flux is zero, with correspondingly zero Yang-Mills energy, so the index is readily interpreted as 
\begin{equation}
\mathcal{I}_{U(N)+B}^{M=0}(\tau| \xi) = \mathcal{I}_{U(1)} \mathcal{I}_{SU(N)/\Z_N}^{\theta=0}(\tau| \xi) = \mathcal{I}_N(\tau|\xi).
\end{equation}


What about the other sectors with $M\ne 0$? Although the bundles with non-zero field strength 
have non-zero Yang-Mills action, these $(M,N)$-string configurations are still half-BPS in the the string theory target space, and must still preserve 16 supersymmetries! Explicitly, the D1-brane worldvolume theory 
has non-linearly realized supersymmetries acting on the $U(1)$ center of mass modes, which are the goldstinos of the spontaneously broken translation symmetry in the presence of the D-branes \cite{Harvey:1996gc,Guralnik:1997sy,Guralnik:1997th}. The action or energy of this flux should be considered as part of the binding energy of the $(M,N)$-string, or as the difference in the central charge of the two BPS sectors of the target-space supersymmetry algebra. The binding energy should be attributed to the DBI action \cite{Taylor:1997dy} in the same sense as the tension of the $N$ D1-branes is, and should be excluded from the vacuum describing the fluctuations of the bound state. 
In particular, we can modify the definition of the elliptic genus to count states that are BPS with respect to the supercharges preserved by the bound state, essentially by shifting the Hamiltonian by the central charge of the superalgebra. The corresponding BPS states are exactly the configurations with fixed electric flux $M$ and minimal energy. Since the $U(1)$ factor is free, the fields that contribute to the index 
are unaffected by this modification. Thus, we obtain the index of the $U(N)$ theory for given sector with $M$ units of electric flux, 
\begin{equation} \mathcal{I}_{U(N)+B}^M(\tau|\xi) = \mathcal{I}_D (\tau|\xi). \end{equation} This strongly suggests that the vacuum describing the massless fluctuations of the $(M,N)$-string is given by the sigma model into $\Sym^D \R^8$.
Moreover, we also construct the index of the $U(N)+B$ theory that sums over each $(M,N)$-string BPS sector, which is naturally refined by the $U(1)$ holonomies of the $B$-field on the spacetime torus $e^{iM\phi} = e^{i M \int_{T^2} B}$ with representations labeled by the F1-string winding number $M$,
\begin{equation}
\mathcal{I}_{U(N)+B} (\tau|\xi) = \sum_{M\in \Z} e^{iM \phi} \mathcal{I}_{U(N)+B}^M(\tau|\xi) =\sum_{M\in \Z} e^{iM \phi} \mathcal{I}_D (\tau|\xi). \label{eq:D1-brane index}
\end{equation}
We note that this D1-brane index is invariant under the S-duality of the Type IIB string, which is generated by exchanging $M$ and $N$ and shifting $M$ by a multiple of $N$, all the while leaving $D$ invariant.
By an S-duality followed by a T-duality on the circle wrapped by the D-string, the $(M,N)$-string  is mapped to $N$ F-strings bound to $M$ D0-branes \cite{Dijkgraaf:1997vv}. Thus, the index \eqref{eq:D1-brane index} is also an index of the $N$ Type IIA F-strings bound to D0-branes. Our result suggests that the world sheet theory of $N$ F-strings bound to $M$ D0-branes in the free string limit $g_s = 0$ is given by the supersymmetric sigma model into $\Sym^D \R^8$, and in particular, $N/D$ F-strings bound to $M/D$ D0-branes behave like free strings.

The paper is organized as follows. In Section \ref{sec:classical vacua}, we 
analyze the structure of topological sectors of $\msym_2$ for $SU(N)$ and $U(N)$ gauge group, as well as the related $SU(N)/\Z_K$ and $U(N)+B$ theories, and determine the moduli space of flat connections and the classical vacua when the spacetime is $T^2$. In Section \ref{sec:SU(N) elliptic genera} we discuss how the elliptic genus generalizes for $SU(N)/\Z_N$ gauge theories 
to include integration over the various components of the moduli space of flat connections. In Section \ref{sec:MSYM elliptic genus}, we compute the elliptic genus of $SU(N)/\Z_K$ $\msym_2$, and infer the elliptic genus for the $U(N)$ theory with and without the $B$ field. Finally, in Section \ref{sec:symmetric orbifold elliptic genus}, we compute the elliptic genus of the $\Sym^N(\R^8)$ sigma model, and establish some of its properties which allow us to match it to the gauge theory elliptic genus. We also include Appendix \ref{apx:action and susy}, which spells out some details about the action and supersymmetry transformations of $\msym_2$.



\section{The structure of vacua}\label{sec:classical vacua}

Bound states of D1-branes with the F-strings in Type IIB string theory suggest that the $\msym_2$ with $SU(N)$ gauge group should have $N$ superselection sectors, and that the full worldvolume theory of the $N$ D1-branes (with $U(N)$ gauge group) should have topological sectors labeled by $\Z$ \cite{Witten:1995im}. A complete description of the vacua of the $\msym_2$ should account for the vacua in these additional sectors as well. Therefore, we will now task ourselves with hunting for them. We will discover that a rich story underlies the various vacua.

\subsection{Topological sectors}

Let's start by focusing on the $SU(N)$ theory. It was shown in \cite{Witten:1995im} that on a worldsheet with boundary, such as $\R^{1,1}$ for concreteness, the sector with $M$ F-strings attached to the stack of $N$ D1-branes manifests itself as a Wilson loop ``at infinity'' in the $M$th tensor power of the fundamental representation of the gauge group. The vacua of superselection sectors of 2d non-abelian theories 
have been analyzed a long time ago by Witten \cite{Witten:1978ka}. Since $\msym_2$ contains only adjoint fields, the center of the gauge group acts trivially on all fields. In particular, the net charge under the center cannot be screened by local fields. For $G=SU(N)$, the center is $Z(G)=\Z_N$. Therefore, we see that the $N$ superselection sectors in the $SU(N)$ theory are labeled by the background $\Z_N$ charge. More precisely, the theory has a $Z(G)$ 1-form global symmetry, for which the charged objects are the Wilson loops in $SU(N)$ representations \cite{Kapustin:2014gua}, and the corresponding conserved $\Z_N$ charge labels the superselection sectors. The creation of a Wilson loop in representation $R$ will act as a domain wall between two superselection sectors of $\Z_N$ charge differing by the charge under the center (or $N$-ality) $N_R$ of the representation. 

We would like to be able to identify and isolate the vacua. This is best done if one declares the gauge group to be $G_{adj}=SU(N)/\Z_N$, which we can do since all the fields are 
uncharged under the $\Z_N$ center. 
Indeed, the $\msym_2$ Lagrangian \eqref{eq:msymlag} with the fields taken to be valued in $\mathfrak{su}(N)$ does not uniquely define a quantum field theory, since one can declare the gauge group to be any Lie group with Lie algebra $\mathfrak{su}(N)$. This choice does not affect the local physics, but determines which non-local operators and instanton sectors are present in the theory. For example, the theory with $SU(N)$ gauge group has Wilson loops in all $SU(N)$ representations, whereas the $SU(N)/\Z_N$ theory only has Wilson loops in representations for which $N_R \equiv 0$, but also has surface operators which have boundary Wilson loops in arbitrary $SU(N)$ representations (we will revisit these surface operators shortly
). Moreover, because $\pi_1(SU(N)/\Z_N) = Z(SU(N))=\Z_N$, the $SU(N)/\Z_N$ theory has a total of $N$ instanton sectors. When the worldsheet is $\R^{1,1}$, the instanton sectors were described in \cite{Witten:1978ka}. More generally, if one considers the $SU(N)/\Z_N$ gauge theory on a closed Riemann surface $\Sigma$, the instanton sectors are the $N$ $SU(N)/\Z_N$-principal bundles on $\Sigma$, labeled by discrete non-abelian 't Hooft electric flux \cite{Hooft:1979uj} --- or, mathematically, the second Stiefel-Whitney class of the bundle \cite{Henningson:2009aa}
\be w_2(P)\in H^2(\Sigma, \pi_1(G_{adj})) =H^2(\Sigma, \Z_N).\ee The $G_{adj}$ theory has additional data in the form of the discrete $\theta$ angle, which takes values in the Pontryagin dual $\Z_N$ of $\pi_1(G_{adj})$. For each of the $N$ choices of the $\theta$ angle, the theory isolates a corresponding superselection sector of the $SU(N)$ theory, and the Hilbert space is a restriction of the $SU(N)$ Hilbert space to that sector. This structure mirrors the structure of vacua in the closely related pure Yang-Mills theories with $SU(N)$ and $SU(N)/\Z_N$ gauge group \cite{Lawrence:2012ua}.

The $SU(N)/\Z_N$ and $SU(N)$ theories are of course closely related. One can obtain the $SU(N)/\Z_N$ theory from the $SU(N)$ theory by gauging the 1-form symmetry generated by the center $\Z_N = Z(SU(N))$ \cite{Kapustin:2014gua}. The procedure is illuminating, as it allows one to explicitly construct the surface operator that detects $w_2$. One can first enhance the $SU(N)$ gauge field to a $U(N)$ gauge field by adding in the trace component $\hat A$, and then impose the $U(1)$ 1-form gauge symmetry generated by
\begin{equation}
\hat A \rightarrow \hat A - N \lambda
\end{equation} which removes the field strength for $\hat A$ and also enhances the allowed gauge bundles to $SU(N)/\Z_N$ bundles. In the resulting $SU(N)/\Z_N$ theory, there are no Wilson loops in representations of $SU(N)$ that transform nontrivially under the center $\Z_N$, unless they are the boundary of a surface operator constructed from $ d\hat A$, which is now a 2-form gauge field. 
The closed surface operator \be e^{i M \int_{\Sigma} {d\hat A}/{N}} \ee evaluates to $e^{i 2\pi M k /N}$ for a bundle with 't Hooft flux $\int_\Sigma w_2 = k$ around the two-cycle represented by $\Sigma$. The integral here is schematic, as $d\hat A$ is not a globally-defined 2-form, instead one should integrate it as a Deligne-Belinson cocycle (see \cite{Kapustin:2014gua} and references therein).\footnote{Heuristically, given a cover $U_i$ of the base, the transition functions $\lambda_{ij}$ on double overlaps and the cocycle conditions on triple overlaps of $d\hat A$ encode the same information as the 't Hooft flux of the $SU(N)/\Z_N$-bundle \cite{Kapustin:2014gua}. The integral extracts that data.}
This operator can be inserted into the path integral to obtain the $SU(N)/\Z_N$ theory with the discrete $\theta$ angle equal to $ 2\pi M / N$. The parameter $M$ is quantized in integer units, as required by invariance under large gauge transformations. 


Even as classical theories, the $G$ theory and the $G_{adj}$ theory are different. In particular, the $G_{adj}$ theory has additional classical field configurations corresponding to connections on $G_{adj}$-bundles, even for those which are not $G$-bundles. Each of these bundles admit flat connections, so the moduli space of classical vacua of Yang-Mills theory on Riemann surfaces is enlarged to include flat connections of $G_{adj}$-bundles on the Riemann surface. For theories with supersymmetry, one expects zero energy field configurations supersymmetric to flat connections for the non-trivial $G_{adj}$-bundles. We will describe these configurations in Section \ref{subsec: classical vacua in instanton sectors}, and find a pleasant parallel to the string theory predictions for the vacua.

It is perhaps good practice to say a few words about the definition of a gauge theory with gauge group $G$ and solidify our footing. In accordance with the literature \cite{Aharony:2013hda}, we take a general $G$-gauge theory to satisfy the following properties:
\begin{enumerate}
\item All local fields are in representations of $G$.
\item Wilson lines in all representations of $G$ are present.
\item The path integral sums over all $G$-bundles. There could be additional data that determines weights for the sum over $G$-bundles.
\end{enumerate} With these properties, the difference between a $G$ and $G/H$ theory where $H\subset Z(G)$ is made explicit. We can go ahead and generalize our above analysis by also defining the $SU(N)/\Z_K$ $\msym_2$ theory with $K|N$ accordingly. The 2d $SU(N)/\Z_K$ theory has $K$ instanton sectors, weighted by a $\Z_K$ valued discrete $\theta$ angle. Since the theory contains only adjoint fields, the charge under the center $\Z_{N/K} = \Z_N/\Z_K$ will not be screened, and for each choice of the $\theta$ angle the theory will have $N/K$ superselection sectors corresponding to those superselection sectors of the $SU(N)$ theory with $\Z_N$ charge congruent modulo $K$ to a given value determined by the choice of $\theta$.

Let's return to the $U(N)$ $\msym_2$. The ``standard'' $U(N)$ $\msym_2$ has superselection sectors analogous to the $SU(N)$ $\msym_2$ The pure $U(N)$ Yang-Mills theory in 2d has $N$ superselection sectors \cite{Lawrence:2012ua}. Similarly, a 2d $U(N)$ gauge theory without fields charged under the center of the gauge group also has $N$ superselection sectors, thus so does $U(N)$ $\msym_2$. The $U(N)$ theory has instanton sectors labeled by the integers corresponding to the quantized electric flux (or vortex number) $c_1\in H^2(\Sigma,\Z)$. Although one might hope to identify these sectors with the $(M,N)$-string sectors, this turns out to be not quite right.  
The true theory describing the interactions of $N$ D1-branes is not just the $U(N)$ $\msym_2$ that we described above by the action \eqref{eq:msymlag}, but also has a 2-form gauge field $B$ coming from the restriction of the Kalb-Ramond field present in the NS-NS sector of the string theory target space to the brane worldvolume. The $B$-field plays a subtle and important role, primarily by enhancing the classical field configurations of the theory. The $B$-field, being a 2-form gauge field, has Abelian 1-form gauge transformations under which the $U(N)$ gauge field $A$ also transforms \cite{Witten:1995im},
\begin{align} B &\rightarrow B + d \lambda, \\ A &\rightarrow A - \lambda \mathbf{1}_N, \end{align} where $\lambda$ is the 1-form gauge transformation parameter and $\mathbf{1}_N$ is the $N\times N$ identity matrix generating the center of the $\mathfrak{u}(N)$ algebra. The correct gauge-invariant Lagrangian has the following kinetic term for the gauge field,
\be -\frac{1}{4} \Tr \; (F_{\mu \nu} + B_{\mu \nu} \mathbf{1}_N)^2, \ee and $\mathcal{F} = F+B  \mathbf{1}_N$ is the appropriately modified field strength. Writing the $U(N)$ gauge field as \be A = \frac{1}{N} \hat A \mathbf{1}_N + A', \ee with $\hat A$ the $U(1)$ gauge field corresponding to the trace and $A'$ the leftover $SU(N)/\Z_N$ gauge field, we note that the 1-form gauge transformation above acts only on the $U(1)$ gauge field $\hat A$.
Since all of the scalar and fermion fields are in the adjoint, $\hat A$ only appears in the gauge field kinetic term in the Lagrangian, and therefore none of the rest of the Lagrangian is modified with the inclusion of the $B$-field, as they are already gauge invariant under the 1-form gauge symmetry. The $\Ncal=(8,8)$ supersymmetry remains intact once one modifies the supersymmetry transformations accordingly by replacing $F$ with $\mathcal{F}$.

Now, let's consider what gauge bundles the theory has. As can be seem from the equation of motion for $\hat A$, $\Tr \mathcal{F}$ is constant, and has periods quantized in integer units when we impose the parameter $\lambda$ generates the gauge group $U(1)$ instead of $\R$ \cite{Witten:1995im}. So the theory considered on a Riemann surface $\Sigma$ has a topological quantum number labeled by $\tilde{c}_1 = [\Tr \mathcal{F}/2\pi] \in H^2(\Sigma,\Z)$ corresponding to the generalized $U(1)$ electric flux. For an honest $U(N)$ theory --- without the $B$-field --- the single Chern class $c_1 = [\Tr F / 2\pi] \in H^2(\Sigma, \Z)$  would classify all $U(N)$-principal bundles. A $U(N)$-bundle can be thought of as the data of a $U(1)$-bundle and an $SU(N)/\Z_N$-bundle, such that the Stiefel-Whitney class of the $SU(N)/\Z_N$-bundle $w_2\in H^2(\Sigma,\Z_N)$ is related to the $U(1)$ characteristic class as $\int w_2 = \int c_1 \pmod{N}$ \cite{Atiyah:1983aa}. This can be seen at the level of the transition functions for the gauge field. However, in the theory with the $B$-field, the additional 1-form symmetry enhances the transition functions and generalizes the allowed bundles and connections, as detailed in \cite{Kapustin:2014gua,Kapustin:1999di}. 
The resulting generalized $U(N)$-connection admits an independent 't Hooft flux $w_2$ in addition to the electric flux $\tilde{c}_1$. 
This type of gauge bundle would be more accurately described in the language of gerbes or 2-bundles, but we will not need to go into such territory here.
Due to the particularly simple 2-group structure, practically speaking we can think of the allowed gauge bundles as $U(1)\times SU(N)/\Z_N$-bundles, with independently chosen characteristic classes $(\tilde{c}_1,w_2) \in H^2(\Sigma,\Z) \times H^2(\Sigma,\Z_N)$. The classical configurations of the scalar and fermion fields in the theory mimic the configurations in a $U(1)\times SU(N)/\Z_N$ theory. It is important to emphasize that the theory is \emph{not} a $U(1)\times SU(N)/\Z_N$ gauge theory; for example the operator content --- such as Wilson lines and surface operators --- is different. 

Configurations with $\int_\Sigma \Tr \mathcal{F}/2\pi = M$ correspond to the binding of $M$ F-strings \cite{Witten:1995im}. The $M$ units of flux is interpreted as the NS-NS charge carried by the F-string, and $\Tr \mathcal{F}$ serves as a source for the $B$-field in the string target space. The generalized Yang-Mills action (or energy) of the flux is the binding energy of the $(M,N)$-string, measured as the difference from the mass of the $N$ D-strings. If one considers the theory on the cylinder $C=\R_t \times S^1$, the presence of $M$ units of $\Tr \mathcal{F}$ flux implies that there is a Wilson loop
\begin{equation}
e^{i M \oint_{\partial C} \hat A /N}
\end{equation} at the boundary. However, this Wilson loop must also be complemented by the $B$-field to be gauge invariant.
This can be seen by noting that the standard $U(N)$ Wilson loops are not gauge invariant in this theory, instead one has the following surface operators considered in \cite{Kapustin:2014gua},
\begin{equation}
\left(\Tr_R \mathcal{P} \exp \oint_{\partial \Sigma'} A \right) e^{i N_R \int_{\Sigma'} B} = \left(\Tr_R \mathcal{P} \exp \oint_{\partial \Sigma'} A' \right) e^{i N_R \int_{\Sigma'} \frac{d\hat A}{N} + B}.
\end{equation} Note that the inside and outside of this operator differ by $N_R$ units of $U(1)$ electric flux $\Tr \mathcal{F}$. 
So, the sector with $M$ units of electric flux has the operator 
\begin{equation}
e^{i M \oint_{\partial C} \frac{\hat A}{ N}} e^{iM\int_C B} = e^{i M \int_{C} \frac{d \hat A}{N} }e^{i M \int_{C} B}
\end{equation} turned on. As with the $SU(N)/\Z_N$ theory, the integral of the 2-form gauge fields $d\hat A$ and $B$ are not of global 2-forms.
Upon quantizing the theory on the cylinder, these states with $M$ units of electric field $\Tr \mathcal{F}$ are the $(M,N)$-string states.
They fall into $N$ superselection sectors determined by $M \pmod{N}$.

We are interested in the low-energy fluctuations of the $(M,N)$-string bound states. The path integral of the worldvolume $U(N) + B$ theory on the Euclidean torus $T^2$ is naturally a trace of the theory quantized on the cylinder $C$. The trace sums over the $(M,N)$-string sectors by summing over the flux $\tilde c_1\in H^2(T^2,\Z)$. 
Crucially, the $U(N)+B$ theory 
has the operator 
\begin{equation}
e^{iM\int_{T^2} d\hat A /N} e^{i M \int_{T^2} B}
\end{equation} turned on in the sector with $M$ units of electric flux. On a closed surface such as $T^2$, the first factor measures the 't Hooft flux in the $SU(N)/\Z_N$ sector, since $\int_{T^2} d\hat A = \int_{T^2} w_2$ exactly as for the $SU(N)/\Z_N$ theory discussed above. Once again, the presence of this term provides a discrete $\theta$ angle $2\pi M/N$ for the sum over the $SU(N)/\Z_N$-bundles. The second factor is simply the Wilson surface operator for the $U(1)$ 1-form gauge symmetry. The ``charge'' $M$ is nothing but the F-string winding number once again. This closed Wilson surface operator measures the $U(1)$-valued holonomy of the background $B$-field.

We note that for the $U(N)$ theory with or without the $B$ field, one can also add a continuous $\theta$-angle term to the action proportional to $\int \Tr \mathcal{F}$ or $\int \Tr F$, or in general a supersymmetric FI parameter. For the theory with the $B$ field, this $\theta$ angle is related to the axion of the Type IIB string theory \cite{Witten:1995im}. However we will not consider including this term, as it does not affect the qualitative features of our discussion (or the elliptic genus).

\subsection{Classical vacua on $T^2$} \label{subsec:Classical vacua on torus}

Motivated to perform a quantitative check of our conjectures regarding the structure and description of the vacua, we would like to compute the elliptic genera of the $\msym_2$ theory with the various gauge groups discussed above. The elliptic genus is a certain supersymmetric partition function on the 2-torus $T^2$ \cite{Witten:1986bf}, which counts (with a sign $(-1)^F$) states in the cohomology of a conjugate pair of right-moving supercharges $\mathcal{Q}_R^\pm$.\footnote{Elliptic genera can be defined for theories with $\Ncal=(0,1)$ supersymmetry as well, with a single self-conjugate right moving supercharge $\mathcal{Q}_R$. However, one expects less control over the spectrum, as generically R- and flavor symmetries can be discrete.} States in the cohomology correspond to right-moving vacua tensored with left-moving BPS states. Elliptic genera have been extensively used to study $\Ncal=(2,2)$ and more recently $\Ncal=(0,2)$ theories; for a very restricted set of examples see \cite{Witten:1993jg,Benini:2013nda,Benini:2013xpa,Gadde:2013lxa}. It is often useful to refine the elliptic genus by other conserved charges in the theory that commute with $\mathcal{Q}_R^\pm$, which allows more information about the spectrum of the theory to be captured. For a theory with at least $\Ncal=(0,2)$ supersymmetry, the elliptic genus can be schematically defined as 
\be \mathcal{I} = \Tr \; (-1)^F \prod_{J_L} y^{J_L} \prod_f x^f q^{H_L} \bar{q}^{H_R},
\ee where $J_L$ stands for the generators of left-moving R-symmetry, and $f$ stands for the generators of bosonic flavor symmetries, all commuting with the $\mathcal{Q}_R^\pm$. With this philosophy, the definition of the elliptic genus can be extended to theories with higher supersymmetry, as we will do so for theories with $\Ncal=(8,8)$ supersymmetry in Sections \ref{sec:MSYM elliptic genus} and \ref{sec:symmetric orbifold elliptic genus}. The trace can be taken in the Ramond or Neveu-Schwartz left- and right-moving Hilbert spaces of the theory on the spatial circle. We will specialize to the Ramond-Ramond sector.
The elliptic genus is invariant under deformations of a theory preserving the right-moving supercharges, and therefore is a topological index of theories. In particular, it is invariant under RG flow, which allows it to be computed in the free UV limit of a theory. For example, for Landau-Ginzburg theories it is sufficient to know the contributions from the field content of the theory in the free limit and impose the restrictions on R- and flavor symmetries coming from the superpotential \cite{Witten:1993jg}.

For gauge theories the elliptic genus can be computed in the free limit of the theory by introducing fugacities for the gauge charges, which amounts to doing the path integral in the presence of a fixed but arbitrary background flat gauge connection, and then imposing Gauss\rq{} Law to project onto physical states by integrating over the moduli space of flat connections \cite{Gadde:2013dda,Benini:2013nda,Benini:2013xpa}. 
As discussed, for gauge theories with only adjoint fields such as $\msym_2$, one has freedom in choosing the global form of the gauge group. For example, the theory with $SU(N)$ gauge group differs from the theory with $SU(N)/\Z_K$ gauge group for any $K|N$, despite having the same Lagrangian. Since $\pi_1(SU(N)/\Z_K) = Z_{K}$, 
the $SU(N)/\Z_K$ theory has additional classical field configurations on $T^2$, therefore both the moduli spaces of flat connections and the moduli space of classical vacua are enhanced to include various disconnected components. 
These additional components are crucial for the computation of the elliptic genus for such theories, as the path integral sums over them as well. We note that to compute the elliptic genus of the $SU(N)$ theory and the $U(N)$ theory without the $B$ field, we only to integrate over the trivial moduli space of the $SU(N)$ bundle. However, to compute the elliptic genus of the $U(N)$ theory with the $B$ field, we need to integrate over the full $SU(N)/\Z_N$ moduli space. Also, once we have a description of the $SU(N)/\Z_N$ moduli space, we can infer the $SU(N)/\Z_K$ moduli space, and compute the elliptic genera for the $SU(N)/\Z_K$ theories for free. To prime ourselves for computing the elliptic genera, we now turn to a description of the moduli space of flat $SU(N)/\Z_N$-connections on $T^2$. As an added bonus, we will be able to understand the classical field configurations on $T^2$ for the various theories discussed, and discover the classical vacua.

\subsubsection{Flat connections on $SU(N)/\Z_N$-bundles over $T^2$}

%
A treatment of the moduli spaces of flat connections for $SU(N)/\Z_N$ bundles was given in \cite{Schweigert:1996tg}, where in particular it was shown that the moduli spaces for the topologically non-trivial bundles with structure group $G$ are isomorphic to moduli spaces of trivial bundles for a different structure group $G^\omega$. Here, we will give a self-contained, very explicit, and somewhat pedestrian account of the moduli spaces of flat connections on $T^2$, specializing to the structure group $G_{adj}=SU(N)/\Z_N$. 

Flat connections can be solved for by their holonomies, and the moduli space is given by
\be \mathcal{M}_{\rm flat} = \Hom(\pi_1 (T^2) , G_{adj} )/{G_{adj}}. \ee
Denoting elements of $SU(N)/\Z_N$ as conjugacy classes $[A]$ of elements $A\in SU(N)$, such homomorphisms for $G_{adj} = SU(N)/\Z_N$ is the set of solutions to the equation 
\be [A] [B] [A]^{-1} [B]^{-1} = 1 \label{eq:holonomy} \ee 
modulo conjugation by $SU(N)/\Z_N$ (or, equivalently, by $SU(N)$ as the center acts trivially). For $SU(N)$, the analogous equation $ A B A^{-1} B^{-1} =1$ implies $A$ and $B$ lie in the same maximal torus. While such commuting holonomies describe flat $SU(N)/\Z_N$ connections, they are not the only solutions to \eqref{eq:holonomy}. To find the rest of the solutions, we can lift \eqref{eq:holonomy} to $SU(N)$, and find solutions there. In $SU(N)$, we have $N$ equations, 
\be A B A^{-1} B^{-1} = \omega_N^k, \label{eq:holonomyLifted} \ee
labeled by $k \in \Z/N \Z$, that project to the equation \eqref{eq:holonomy} in $SU(N)/\Z_N$. In \eqref{eq:holonomyLifted}, $A$ and $B$ are now in $SU(N)$ and $\omega_N$ is a primitive $N${th} root of unity. We can use part of the gauge freedom to diagonalize $B$, leaving only the Weyl group, which reorders the eigenvalues. The equation now reads 
\be S D S^\dagger = \omega_N^k D, \ee
which is an eigenvalue equation for conjugacy action of $SU(N)$ on a diagonal matrix. For each $N$ and $k$, there is always a solution, constructed from the clock and shift matrices\footnote{We note that as defined, $D_N$ and $S_N$ do not always have determinant equal to $1$, and therefore are not always in $SU(N)$. This can easily be fixed by dividing by the $N$th root of the determinant in the definition. Since this overall phase decouples from the conjugation action, and so does not affect our calculations, we will drop it to avoid clutter.}
\be D_N =\begin{pmatrix} 1 & & & \\ &  \omega_N & & \\ & & \ddots & \\ & & & \omega_N^{N-1} \end{pmatrix} ,  \text{ and} \quad S_N = \begin{pmatrix*}[l] 0 \; &1 \; & & & \\ &  0 & 1 & & \\ & & & \ddots \; &1 \\ 1 & & & & 0 \end{pmatrix*} \ee 
which satisfy 
\be S_N^k D_N (S_N^k)^\dagger = \omega_N^k D_N. \ee Correspondingly, the pair of holonomies $([S_N^k],[D_N])$ describes a flat $SU(N)/\Z_N$ connection. Therefore, each $k$ contributes a new component, $\Mcal_{N,k}$, to the moduli space of flat $SU(N)/\Z_N$ connections, $\Mcal_N$. These components are disjoint, and labeled by discrete data $k$, so we can write 
\be \Mcal_N = \bigsqcup_{k=0}^{N-1} \Mcal_{N,k}. \ee
The principal $SU(N)/\Z_N$ bundle $P_{N,k}$ on $T^2$, with 't Hooft non-abelian flux $k=\int_{T^2} w_2(P_{N,k})$, has the moduli space of flat connection precisely $\Mcal_{N,k}$. 

Let\rq{}s proceed to describe $\Mcal_{N,k}$ for given $N$ and $k$. It will be useful to define $d=\gcd(N,k)$, as $\Mcal_{N,k}$ will turn out to have complex dimension $d-1$. In fact, for given $N$ and any two $k_1$ and $k_2$ such that $d=\gcd(N,k_1) = \gcd(N,k_2)$, we will have the isomorphism $\Mcal_{N,k_1}\cong \Mcal_{N,k_2}$. This is not a surprise, since the bundles $P_{N,k_1}$ and $P_{N,k_2}$ are related by an automorphism of $\pi_1(SU(N)/\Z_N) = \Z_N$ exchanging $k_1$ and $k_2$. Motivated by this, we define $\Mcal_{N,d}\cong \Mcal_{N,k}$. Let\rq{}s start with the case when $N$ and $k$ are relatively prime, so $d=1$. 

\paragraph{Moduli space of bundles with {$d=1$}:}

We first note that for any pair of elements $(A,B)$ in $SU(N)$ satisfying some commutation relation, such as \eqref{eq:holonomyLifted}, there are a total of $N^2$ points $(\omega_N^a A, \omega_N^b B)$, where $a,b = 1,2,\dots,N$, that do so. (This is necessary for $SU(N)$ solutions $(A,B)$ to descend to $SU(N)/\Z_N$ solutions $([A],[B])$.) So, we can work with representatives $(A,B)$ of the conjugacy class $([A],[B])$.

To solve \eqref{eq:holonomyLifted}, we can diagonalize either $A$ or $B$, and obtain the solutions $(D_N^m,S_N^n)$ or $(S_N^m,D_N^n)$, for some $mn=k$. We note that $S_N$ generates the $\Z_N$ subgroup of the Weyl group $\mathbb{S}_N$, and therefore has the same eigenvalues as $D_N$ (up to an irrelevant determinant factor). So, 
$S_N$ and $D_N$ are conjugate and the solutions $(S_N^m,D_N^n)$ and $(D_N^m,S_N^n)$ are identified by gauge transformations. Also, since we necessarily have $\gcd(m,N) = \gcd(n,N) =1$, the solutions for various $m,n$ only reorder the eigenvalues of $D_N$ and $S_N$ up to an overall cyclic ordering, and are related by the action of the Weyl group
. We can partially fix the gauge by choosing $m=k$ and $n=1$, and we are left with $N^2$ solutions in $SU(N)$ given by $(\omega_N^a S_N^k, \omega_N^b D_N)$. But, precisely because $S_N D_N S_N^\dagger = \omega_N D_N$, these $N^2$ points are also identified by gauge transformations generated by the simultaneous conjugation by $D_N$ and by $S_N$,
\begin{align}
S_N(\omega_N^a S_N^k, \omega_N^b D_N)S_N^\dagger &=(\omega_N^a S_N S_N^k S_N^\dagger, \omega_N^b S_N D_N S_N^\dagger) = (\omega_N^a S_N^k, \omega_N^{b+1} D_N)
\\ D_N(\omega_N^a S_N^k, \omega_N^b D_N)D_N^\dagger &= (\omega_N^a D_N S_N^k D_N^\dagger, \omega_N^b D_N D_N D_N^\dagger) = (\omega_N^{a-1} S_N^k, \omega_N^{b} D_N)
\end{align} so there is a single solution in $SU(N)$ up to conjugacy. Projecting to $SU(N)/\Z_N$, we still have a single point, $([S_N^k],[D_N])$, of $SU(N)/\Z_N$ holonomies, but this point is fixed at order $N^2$ by the $\Z_N^2$ generated by simultaneous conjugation by $[D_N]$ and by $[S_N]$, 
\begin{align} [S_N] ([S_N^k],[D_N]) [S_N]^\dagger  &= ([S_N^k],[D_N]) 
\\  [D_N] ([S_N^k],[D_N]) [D_N]^\dagger  &= ([S_N^k],[D_N]) \end{align}
So, we finally have
\be \Mcal_{N,k}= \{([S_N^k],[D_N])\} / \Z_N^2.\ee We see that \be \Mcal_{N,d=1} =\{([S_N],[D_N])\} / \Z_N^2, \ee and the isomorphism $\Mcal_{N,d=1}\cong  \Mcal_{N,k}$ is given by replacing the primitive $N$th root of unity $\omega_N$ by its $k$th power.

\paragraph{Moduli space of bundles with {$d>1$}:}

The essential observation for the $d\ne1$ cases is that 
\be S_N^d = S_{N/d}\otimes I_d, \text{ and } D_N^d = D_{N/d}\otimes D_d^{d/N}. \ee
Since the $d$-dimensional factors commute, one can turn on arbitrary eigenvalues in the corresponding $d$-dimensional subgroup of the Cartan torus.
Explicitly, the solutions are generalized to 
\be (e^{i\mathfrak{h}_{N,d} (\theta_s)} S_N^k,e^{i\mathfrak{h}_{N,d}(\theta_t)} D_N)=(S_{N/d}\otimes e^{i\mathfrak{h}_{d} (\theta_s)},D_{N/d}\otimes e^{i\mathfrak{h}_{d} (\theta_t)} D_{N}^d), \ee
where
\be e^{i\mathfrak{h}_{N,d}(\theta)} := I_{N/d} \otimes e^{i\mathfrak{h}_{d} (\theta)} :=  I_{N/d} \otimes \begin{pmatrix} e^{2 \pi i\theta_1} & & & \\ &   e^{2 \pi i\theta_2} & & \\ & & \ddots & \\ & & &  e^{2 \pi i\theta_d} \end{pmatrix},\ee as one can easily check that
\be
\begin{split} (e^{i\mathfrak{h}_{N,d} (\theta_s)} S_N^k) (e^{i\mathfrak{h}_{N,d}(\theta_t)} D_N ) & (e^{i\mathfrak{h}_{N,d} (\theta_s)} S_N^k)^\dagger 
\\ &= S_{N/d}^{k/d} D_{N/d} (S_{N/d}^{k/d})^\dagger  \otimes e^{i\mathfrak{h}_{d} (\theta_s)} (e^{i\mathfrak{h}_{d} (\theta_t)} D_d^{d/N}) e^{-i\mathfrak{h}_{d} (\theta_s)}
\\ &= \omega_{N/d}^{k/d} D_{N/d} \otimes e^{i\mathfrak{h}_{d} (\theta_t)} D_d^{d/N}
\\ &= \omega_N^k( e^{i\mathfrak{h}_{N,d}(\theta_t)} D_N).
\end{split}
\ee
The unitarity condition fixes $(\theta_s)_i$ and $(\theta_t)_i$ to be real, and the determinant condition fixes their sums to zero. Assigning the two holonomies to the spatial (along $1$) and temporal (along $\tau$) directions of the base torus, the moduli space inherits a natural complex structure, and is parametrized by complex coordinates $u_i = (\theta_t)_i - \tau (\theta_s)_i$ which are periodic: $u_i \sim u_i +1 \sim u_i +\tau$.

In choosing this presentation of the holonomies, we have used part of the gauge symmetry to write them as products of factors of size $N/d$ and $d$. 
We are left with a $\mathbb{Z}_{N/d}^2\times \mathbb{S}_{d} $ subgroup of the gauge group. To see this, note that as far as the $N/d$ by $N/d$ factor is concerned, the situation is analogous to the $d=1$ case, wherein we have used part of the gauge symmetry to order the eigenvalues of $S_{N/d}$ and $D_{N/d}$ up to a cyclic ordering, and there is a remaining $\Z_{N/d}^2$, generated by simultaneous conjugation by $S_{N/d}\otimes I_d$ and by $D_{N/d}\otimes I_d$, corresponding to the cyclic reordering of the eigenvalues, which acts on the solutions by identifying $u_i \sim u_i +\frac{1}{N/d} \sim u_i +\frac{\tau}{N/d}$. The $d \times d$ block also has its eigenvalues permuted by the Weyl group $\mathbb{S}_d$ of the $d$-dimensional Cartan subgroup. So, in $SU(N)$, the moduli space is $\tilde{\mathfrak{M}}_{N,k}/\mathbb{S}_d$ where
\begin{equation}
\tilde{\mathfrak{M}}_{N,k} := \left\{( S_{N/d}^{k/d} \otimes e^{i\mathfrak{h}_{d} (\theta_s \cdot N/d)}, D_{N/d}\otimes e^{i\mathfrak{h}_{d}(\theta_t \cdot N/d)} )\right\} \cong (T^2/\Z_{N/d}^2)^{d-1}.
\end{equation} Here, $T^2$ is a copy of the base torus, with the same complex structure.

Once we project to $SU(N)/\Z_N$, the coordinates undergo the further identifications, $u_i \sim u_i +\frac{1}{N} \sim u_i +\frac{\tau}{N}$, so the solutions are fixed by the $\Z_{N/d}^2$ action above. The moduli space is then
\be \Mcal_{N,k} \cong \{ ([ S_{N/d}^{k/d} \otimes e^{i\mathfrak{h}_{d} (N \theta_s)}], [D_{N/d} \otimes e^{i\mathfrak{h}_{d} (N \theta_t)}] ) \} /  \Z_{N/d}^2 \times \mathbb{S}_d 
. \ee 
Once again, dependence on $k$ is only through $d$, via the choice of an $N/d$th root of unity, and we can define 
\be \Mcal_{N,d} \cong (\mathfrak{M}_{N,d}/\mathbb{S}_d)/\Z_{N/d}^2 ,\ee
where \be \mathfrak{M}_{N,d} =  \left\{( S_{N/d} \otimes e^{i\mathfrak{h}_{d} (N \theta_s)}, D_{N/d}\otimes e^{i\mathfrak{h}_{d}(N \theta_t)} )\right\} \cong (T^2/\Z_N^2)^{d-1},  \label{eq:MfrakNd}\ee
and analogously for its lift to $SU(N)$ via $\tilde{\mathfrak{M}}_{N,d} \cong \tilde{\mathfrak{M}}_{N,k}$. Note that $\Mcal_{N,d} \cong \Mcal_{N/d,1}\times\Mcal_{d,d}$.



\subsubsection{Classical vacua in instanton sectors} \label{subsec: classical vacua in instanton sectors}

The classical zero energy configurations in the $SU(N)/\Z_N$ theory are gauge invariant solutions to the BPS equations,
\begin{equation} \begin{split} F_{\mu\nu} &=0, 
\\ [X^i,X^j]&=0, \\ D_\mu X^i &=0, \end{split}
\end{equation} as can be seen from the fermionic supersymmetry variations, or directly from the action. In the IR limit as $g\rightarrow \infty$, we can think of a particular solution as the data of a flat connection $A_\mu$, and commuting constant bosons $X^i$ satisfying $[A_\mu, X^i]=0$. 
In the sector with trivial instanton number $k=0$, the two components of $A$ commute, so the $X^i$ are all in the same Cartan subalgebra $\mathfrak{h}$, with the Weyl group $W$ permuting the eigenvalues, so the eigenvalues parametrize $(\mathfrak{h})^8/W = (\mathbb{R}^8)^{N-1}/\mathbb{S}_N$ \cite{Dijkgraaf:1997vv}.
However, in the presence of flat connections for non-trivial bundles, zero-energy configurations of the bosons are restricted further. To see directly from the above descriptions of the flat connections which $X^i$ are zero energy, we can exponentiate the relation $[A_\mu, X^i]=0$ to the holonomies of $A_\mu$ as $e^{i\oint A} X^i e^{-i\oint A} =X^i$ for each of the two 1-cycles, the solutions to which are of the form \eqref{eq:classical vacua in instanton sector d}, parametrizing $(\mathbb{R}^8)^{d-1}/\mathbb{S}_d$ for the instanton sector with $d=\gcd(k,N)$. 

For the $U(N)$ theory with the $B$ field, classical field configurations are determined by picking $(\tilde{c}_1,w_2)$, which specifies a gauge 2-bundle. 
Given $(\tilde{c}_1,w_2)$, there will be minimal action configurations with constant field strength $\mathcal{F}_{09}=2\pi \frac{M}{N} \mathbf{1}_N $ and action proportional to $M^2$, where $M=\int \tilde{c}_1$, with the scalars parametrizing $\Sym^d \R^8$. 
(In the Lorentzian theory, such configurations have $M$ units of constant electric flux and energy $g^2 M^2/N$.)
The naive ``zero-energy'' vacua have $\tilde{c}_1= 0$, but, like the $SU(N)/\Z_N$ theory, there are $N$ disconnected components labeled by $w_2$. 

What about the other choices for $\tilde{c}_1$? In the brane picture, the $U(N) +B$ $\msym$ theory is the leading approximation to the brane effective action. One can identify the energy of the flux $\Tr \mathcal{F}$ as the binding energy of F-strings to the D-strings \cite{Witten:1995im,Dijkgraaf:1997vv,Taylor:1997dy}. 
These configurations are half-BPS in the target space, so the corresponding state in the $\msym$ theory should also preserve $16$ supercharges. This is indeed the case, as the $U(N)+B$ $\msym_2$ theory has nonlinearly realized supersymmetries, which are the goldstinos of the breaking of translation symmetry in the presence of D-branes \cite{Harvey:1996gc,Guralnik:1997sy,Guralnik:1997th}, so the supersymmetry variation \eqref{eq:10d susy} of the fermions is corrected to
\begin{equation} \delta \Theta = \Gamma^{MN} \mathcal{F}_{MN} \epsilon_1+ \mathbf{1}_N  \epsilon_2. \end{equation} Here, $\Theta=(\chi,0)^T$ is the 10d Majorana spinor and $\mathbf{1}_N$ is the generator of the center of the $\mathfrak{u}(N)$ algebra. 
In particular, the BPS equations are generalized to 
\begin{equation} \begin{split}
\mathcal{F}_{09}&= \Lambda \mathbf{1}_N, 
\\ [X^i,X^j]&=0, \\ D_\mu X^i &=0, \end{split}
\end{equation} by choosing $\epsilon_2 = -2 \Lambda \Gamma^{09} \epsilon_1$. So, there are BPS sectors with constant $\mathcal{F}_{09}=2\pi \frac{M}{N} \mathbf{1}_N $ such that the minimal action configurations discussed above  --- with constant, commuting $X^i$ parametrizing $\Sym^d(\R^8)$, which are the solutions to the BPS equations for given bundle with $w_2$ --- preserve $16$ appropriately chosen supersymmetries. Therefore, these configurations are ``supersymmetric vacua'', but in a sector with a different central charge of the superalgebra.

We comment that it would be interesting to pursue the relation between the existence of the nonlinear supersymmetry to the presence of the $B$ field.


\section{Elliptic genera of $SU(N)/\Z_N$ gauge theories} \label{sec:SU(N) elliptic genera}

We now delve into the task set upon in \ref{subsec:Classical vacua on torus} of generalizing the elliptic genus when there are additional bundles to consider, such as for $SU(N)/\Z_N$ theories, or for the $U(N)$ theory with the $B$-field.
Once again, as explored in \cite{Gadde:2013dda,Benini:2013nda,Benini:2013xpa}, the elliptic genera of 2d gauge theories is a certain path integral on the torus, which due to localization can be calculated by integrating over the moduli space of flat connections. 
Let $\tilde G$ be a simply-connected semi-simple Lie group with a discrete center $Z(\tilde G)$.
As discussed in the previous section, when one has a Lagrangian with gauge symmetry $\tilde G$ and with all fields invariant under some subgroup $H'$ of $Z(\tilde G)$, one has several distinct choices of theories corresponding to a choice of the global form of the gauge group $G = \tilde G/H$, for each $H\subset H'$. 
These theories will generically have different choices of gauge bundles on the spacetime, and thus the choice of the gauge group will determine which bundles are being summed over by the path integral \cite{Aharony:2013hda}.
For such 2d theories, the elliptic genus is naturally also a sum over the path integrals for the sectors with different gauge bundles, each of which localizes to an integral over the moduli space of flat connections for that bundle.  Furthermore, since $\pi_1(G)= H$, each 2d $G$-gauge theory carries additional discrete data in the form of a $\theta$ angle dual to the relevant characteristic class $w(P)$ of the bundle $P$, which specifies a weight for the sum over components.
So, the elliptic genus can be written schematically as
\be \mathcal{I}^\theta = \sum_{P} e^{i\theta \int w(P)} Z_P, \ee where $Z_P$ is the result of the path integral for the sector of the gauge theory with gauge bundle $P$.  

Concretely, for 2d $SU(N)/\Z_N$ theories, there are $N$ $SU(N)/\Z_N$-bundles $P_{N,k}$, and the relevant characteristic class is $w_2(P) \in H^2(T^2,\Z_N)$, 
with $k=\int_{T^2} w_2(P_k)$, so we write
\be \mathcal{I}_{SU(N)/\Z_N}^\theta = \sum_{k=0}^{N-1} e^{i\theta k} Z_{N,k} \label{eq:SU(N) index path integral}
, \ee where $\theta$ takes values in \be \theta = 0, 2\pi \frac{1}{N},2\pi \frac{2}{N},\dots, 2\pi \frac{N-1}{N}.\ee 

For $U(N)$ $\msym_2$, there is an analogous but slightly more nuanced story. For a standard $U(N)$ theory without the $B$-field, the gauge bundles are $U(N)$-bundles, which are classified by a single integer characteristic class $c_1\in H^2(\Sigma,\Z)$. Only the trivial bundle with $c_1 = 0$ admits flat connections. Since the $U(1)$ degrees of freedom are free and therefore decouple, the elliptic genus is computed as \begin{equation}
\mathcal{I}_{U(N)} = \mathcal{I}_{U(1)} \mathcal{I}_{SU(N)}.
\end{equation}

For the $U(N)$ theory with the 2-form gauge field $B$, recall that there are additional field configurations corresponding to connections on gauge bundles with $G_{adj} = U(1) \times SU(N)/\Z_N$ structure group. On a Riemann surface, these bundles are characterized by two independent classes, $(\tilde{c}_1,w_2)$, however, only certain bundles will contribute to the elliptic genus. For the theory taken at face value, flat connections are only present when $\Tr \mathcal{F}=0$, but there are still the $SU(N)/\Z_N$-bundles with flat connections to sum over, so we have the index
\begin{equation}
\mathcal{I}_{U(N)+B}^{\tilde{c}_1=0} = \mathcal{I}_{U(1)} \mathcal{I}_{SU(N)/\Z_N}^{\theta=0}.
\end{equation}

Let's consider the other sectors, which require adding to the path integral the operator
\begin{equation}
e^{i M \int_{\Sigma} \frac{d \hat A}{N} + B}.
\end{equation} 
As we discussed in Section \ref{sec:classical vacua}, this operator turns on a $U(1)$ electric flux of $M$ units, so we are in the sector with $\tilde{c}_1 = M$.
For the $SU(N)/\Z_N$ sector, $w_2$ is unfixed, and is summed over with the discrete theta angle $\theta = 2\pi M /N$ specified by the operator $e^{i M \int_{\Sigma} \frac{d \hat A}{N}}$. The definition of the elliptic genus for the sector with $M$ strings needs to be modified to take into account the non-linear supersymmetries, which shifts the central charge in the superalgebra. The corresponding elliptic genus localizes to states that saturate the BPS bound in this sector, $\mathcal{F} = 2\pi \frac{M}{N}\mathbf{1} \omega$ with $\omega$ the volume form, 
which specifies the bundle with $\tilde{c}_1 = M$. The scalar and fermionic fields in the $U(1)$ multiplet, as well as the $SU(N)/\Z_N$ sector of the theory are unaffected by this modification. Isolating the holonomy of the $B$-field $e^{i\phi}=e^{i\int_\Sigma B}$, we see that the elliptic genus of this sector is
\begin{equation} e^{iM \phi } \mathcal{I}_{U(N)+B}^M = e^{iM \phi} \mathcal{I}_{1} \mathcal{I}_{SU(N)}^{\theta = 2\pi M/N}, \end{equation} where $\mathcal{I}_{1}$ is the contribution of the free center of mass modes.
As a check, note that for the $U(1)$ theory, the sector with $M$ strings attached, which is the $(M,1)$-string, has index $\mathcal{I}_1$. The S-dual $(1,M)$ string indeed has the same index, if $\mathcal{I}_{SU(N)}^\theta = 1$ for $\theta = 2\pi /N$ which we will show to be the case. 
We can also construct the elliptic genus that sums over each BPS sector (labeled by the $M$ units of flux),
\begin{equation}
\mathcal{I}_{U(N)+B} = \sum_{M\in \Z} e^{iM \phi} \mathcal{I}_{U(N)+B}^M.
\end{equation}

To obtain each of the various indices, the crucial object we need to compute is $\mathcal{I}_{SU(N)/\Z_N}^\theta$. The computation requires some discussion, which we will now elaborate.

\subsection{Integration over components of the moduli space of flat $SU(N)/\Z_N$-connections}\label{subsec:integral over mflat}

To compute $\mathcal{I}_{SU(N)/\Z_N}^\theta$, we need to calculate the path integrals $Z_{N,k}$ for the $SU(N)/\Z_N$ bundles, so let\rq{}s analyze them. In general, $Z_P$ is the path integral over all connections for $P$, so we can write \be Z_P = \frac{1}{\mbox{Vol} (\mathcal{G}(P))} \int_{A\in\Omega^1(T^2,{\rm ad}_P)} \mathcal{D}A \; Z(A). \ee Here, $Z(A)$ is the result of the path integral over all other fields in the presence of a $P$ connection $A$, and $\mathcal{G}(P)$ is the group of gauge transformations (automorphisms) of the bundle $P$. The path integral for the elliptic genus localizes to a finite dimensional integral over the flat connections for the bundle $P$, but there are some global factors we need to worry about.

Let\rq{}s consider the case when the moduli space of flat connections $\Mcal_P$ for a given bundle $P$ is a point. After localization, there are no moduli to integrate over, so the path integral just becomes an evaluation of the torus partition function, $Z_{\rm 1-loop}(u)$, of the fields in the theory in the background of the unique flat connection $u\in\Mcal_{N,1}$ (for a similar example, see the Abelian example in \cite[\S 4.5]{Benini:2013nda}). If the point $u$ is fixed by some finite group of gauge transformations, as is the case for $u\in\Mcal_{N,1}=\mathfrak{M}_{N,1}/\Z_{N}^2$, we should divide by the order of this group. The bundles $P_{N,k}$ with $k\perp N$ (so $d=1$) are exactly of this type, and contribute $Z_{N,k} = Z_{N,1}$ each, with \be Z_{N,1} =  \left. Z_{\rm 1-loop}(u) \right|_{u\in \Mcal_{N,1}} = \frac{1}{N^2} \left. Z_{\rm 1-loop}(u) \right|_{u\in \mathfrak{M}_{N,1}}. \label{eq:massive bundle path integral}\ee

Next, let\rq{}s consider the integral over the trivial $SU(N)/\Z_N$-bundle, $P_{N,k=0}$. Since the bundle $P_{N,0}$ lifts to the (necessarily trivial) $SU(N)$-bundle $\tilde{P}_N$, we can lift the path integral over the $SU(N)/\Z_N$-connection to a path integral $\tilde Z_{\tilde P}$ over an $SU(N)$-connection, $\tilde A$. As analyzed in \cite[\S 4.1]{Witten:1992xu}, the two path integrals are related by a factor of the ratio of the volume of gauge transformations of the bundles, which can be computed using the $N:1$ covering map $\tilde{A}\rightarrow A$ to be 
\be \frac{ {\rm Vol} ( {\mathcal{G}}(\tilde{P}_N) ) }{ {\rm Vol} (\mathcal{G}(P_{N,0})) } = |\pi_1(SU(N)/\Z_N)|^{1-2g} \ee 
on a Riemann surface of genus $g$. Now, the $SU(N)$ path integral is precisely what was shown in \cite{Benini:2013xpa} to localize to a contour integral over the moduli space of flat $SU(N)$-connections, $\tilde\Mcal_{N} = \tilde{\mathfrak{M}}_N/\mathbb{S}_N$. Therefore,
\begin{align}
Z_{N,0} = \frac{1}{N} \tilde Z_{\tilde P_N} = \frac{1}{N}\frac{1}{|\mathbb{S}_N|} \oint_{\tilde{\mathfrak{M}}_N} Z_{\rm 1-loop}. \label{eq:trivial bundle integral lift to SU(N)}
\end{align} The contour integral is determined by the Jeffrey-Kirwan residue operation $\JKRes$. The integrand is once again $ Z_{\rm 1-loop}(u)$, which is naturally a meromorphic function on the $SU(N)/\Z_N$ moduli space $\mathfrak{M}_{N,0}$ for a theory with no fields charged under the center.
Since the $SU(N)$ moduli space $\tilde\Mcal_{N}$ is an $N^2:1$ cover of the moduli space $\mathcal{M}_{N,0} = \mathfrak{M}_{N,0}/\mathbb{S}_N$ of $P_{N,0}$, $Z_{\rm 1-loop}$ extends to a periodic function on $\tilde{\mathfrak{M}}_{N}$. The contours specified by the $\JKRes$ operation only depend on the charges of the fields giving rise to the poles, so the contours on the $SU(N)$ moduli space are also periodic on the $SU(N)/\Z_N$ moduli space for a theory with no fields charged under the center. In particular, the contour integral over $\tilde{\mathfrak{M}}_N$ is just $N^2$ times the contour integral on $\mathfrak{M}_{N,0}$. So, \eqref{eq:trivial bundle integral lift to SU(N)} can be simplified as  \be Z_{N,0} = N \frac{1}{|\mathbb{S}_N|} \oint_{\mathfrak{M}_{N,0}} Z_{\rm 1-loop}. \label{eq:trivial bundle path integral}\ee

Finally, let\rq{}s consider the case with general $k \not \perp N$, so $d>1$. The moduli space in this case is $\Mcal_{N,k} \cong \Mcal_{N,d} = \Mcal_{N/d,1}\times\Mcal_{d,d}$, so flat connections are of the form $A_\mu = (A_{N/d} \otimes A_{d})_\mu$, with $A_{N/d}$ the unique gauge-invariant flat connection on the bundle $P_{N/d,k/d}$, and $A_d$ a flat connection on the bundle $P_{d,0}$ which needs to be integrated over. Combining our arguments above leading to the formulas \eqref{eq:massive bundle path integral} and \eqref{eq:trivial bundle path integral}, the path integral for such $P_{N,k}$ localizes to $Z_{N,k} = Z_{N,d}$ with 
\be Z_{N,d} =\frac{1}{(N/d)^2}  d \frac{1}{|\mathbb{S}_d|} \oint_{\mathfrak{M}_{N,d}} Z_{\rm 1-loop}, \label{eq:general bundle path integral}\ee where $\mathfrak{M}_{N,d}$ as given in \eqref{eq:MfrakNd}. Once again, the contour is determined by the $\JKRes$ operation.

Collecting our results in equations \eqref{eq:massive bundle path integral} and \eqref{eq:general bundle path integral}, the elliptic genus \eqref{eq:SU(N) index path integral} is computed by the formula
\ba \mathcal{I}_{SU(N)/\Z_N}^\theta &=  \sum_{k=0}^{N-1} e^{i\theta k} \gcd(N,k)\frac{1}{|W_{N,k}|} \oint_{\mathfrak{M}_{N,k}}  Z_{\rm 1-loop}(u)
\\ & = \sum_{k \not \perp N} e^{i\theta k} \gcd(N,k) \frac{1}{|W_{N,k}|}  \sum_{u_*\in \mathfrak{M}_{N, d}^*} \JKRes_{u=u_*}(\mathsf{Q}(u_*),\eta) Z_{\rm 1-loop}(u)\nonumber
\\ &\qquad  +   \sum_{k\perp N} e^{i\theta k}\frac{1}{|W_{N,k}|} \sum_{u\in \mathfrak{M}_{N,1}} Z_{\rm 1-loop}(u) \ea with $W_{N,k} = \Z_{N/d}^2 \times \mathbb{S}_d$. We will elaborate on the residue prescription $\JKRes$ in Section \ref{subsec:JKRes on trivial bundle} as part of the computation of the elliptic genus for $\msym_2$.


\subsection{Adjoint fields in the presence of background flat connections}\label{subsec:charge of adjoint fields}

To evaluate the contribution to the index from each of the components of the moduli space, we need to analyze how fields behave in the presence of background flat connections, and determine what $Z_{\rm 1-loop}(u)$ is for each component. 
In line with our end goal, here we will determine $Z_{\rm 1-loop}(u)$ for a theory with all fields in the adjoint representation. 

First off, as is well known, background flat connections on $T^2$ can be interchanged with boundary conditions around the two 1-cycles for fields charged under them. As a simple example, one could keep in mind that the choice of periodic or antiperiodic boundary conditions for fermions is equivalent to the choice of a background flat $\Z_2$-connection. 
Correspondingly, the boundary conditions determine the mode expansions of the fields into oscillators. Since the elliptic genus can be computed in the free field limit, the moding in the presence of arbitrary background flat connections can be easily determined by the charges of the fields.

Let\rq{}s start by considering adjoint fields in the presence of a flat connection for the bundle $P_{N,1}$ over $T^2$, described by a pair of $SU(N)/\Z_N$ holonomies $([S_N],[D_N])$. Although the two matrices $S_N$ and $D_N$ do not commute, their actions by conjugation on $N\times N$ matrices commute, since 
\be S_N D_N A (S_N D_N)^\dagger = \omega_N D_N S_N A (\omega_N D_N S_N)^\dagger = D_N S_N A ( D_N S_N)^\dagger.  \ee
Therefore, the matrices $S_N$ and $D_N$ acting on the Lie algebra $\mathfrak{su}(N)$ by conjugation furnish an $(N^2-1)$-dimensional representation of $\Z_N\times \Z_N$, with eigenvalues $(\omega_N^a,\omega_N^b)$, where $a,b = 0,1,\dots,N-1$, and $a$ and $b$ both not $0$ (as the mode with $a=b=0$ corresponds to the identity matrix, which is not in $\mathfrak{su}(N)$). Explicitly, the eigenspace of the eigenvalue $(\omega_N^a,\omega_N^b)$ is the 1-dimensional vector space of scalar multiples of the matrix $S_N^{-b} D_N^{a}$.
For such a flat connection, adjoint fields will have gauge fugacities $ \exp 2\pi i \frac{a+(-1)^a b \tau }{N} = \omega_N^a q^\frac{(-1)^a  b}{N}$, where the charges $a,b$ are taken from the set
\be \mathfrak{C}_N= \begin{cases}  \{-\frac{N-1}{2},-\frac{N-1}{2}+1,\dots,\frac{N-1}{2}\} &\quad  \text{ for $N$ odd,}
	\\  \{-\frac{N}{2},-\frac{N}{2}+1,\dots,\frac{N}{2},\frac{N}{2}+1\} &\quad  \text{ for $N$ even,} \end{cases} \label{eq:twistedHolonomyCharges}  \ee 
but with the eigenvalue $a=b=0$ excluded. We had to be careful in picking the sign of the exponent of $q$, since we would like our expression to be charge conjugation invariant. This will be necessary later for evaluating the elliptic genus, which is a trace in the Ramond sector. These choices are also invariant under the modular S transformation of the base torus, which amounts to exchanging $a$ and $b$. To summarize, if the contribution to the path integral of modes with gauge fugacity $z=e^{2\pi i u}$ is $\Xi(u)$, the evaluation in \eqref{eq:massive bundle path integral} of $Z_{\rm 1-loop}(u)$ at $u\in \Mcal_{N,1}$ is 
\be \left. Z_{\rm 1-loop}(u) \right|_{u\in \Mcal_{N,1}} = \frac{1}{N^2} \prod_{{\substack{a,b\in \mathfrak{C}_{N} \\ (a,b)\ne(0,0)}}} \Xi\left(\tfrac{a+(-1)^a b \tau }{N}\right). \label{eq:McalN1contribution} \ee The result is identical for all bundles $P_{N,k}$ with $k\perp N$; although the holonomies change to $([S_N^k],[D_N])$, the action on the Lie algebra is isomorphic --- as expected, since they have isomorphic moduli spaces.

Next, we should consider the bundles with moduli spaces of positive dimension. We can study the holonomies $([S_{N/d}\otimes e^{i\mathfrak{h}(\theta_s)}], [D_{N/d}\otimes e^{i\mathfrak{h}(\theta_t)}]) \in \Mcal_{N,d}$, and the result will be the same for all $k$ with $\gcd(k,N)=d$. Similar to our above discussion, conjugation by $S_{N/d}$ and $D_{N/d}$ furnish $d^2$ copies of a $(N/d)^2$-dimensional representation of $\Z_{N/d}^2$. Each of the $d^2$ copies has the usual gauge charges for the adjoint representation of $SU(d)$. Explicitly, the matrices 
\be S_{N/d}^a D_{N/d}^b \otimes (E_{(d)})_{i,j} \ee diagonalize the conjugation action, with eigenvalue
\be \left(\omega_{N/d}^b \; e^{2\pi i ((\theta_s)_i-(\theta_s)_j)}, \omega_{N/d}^{-a} \; e^{2\pi i ((\theta_t)_i-(\theta_t)_j)}\right),  \ee
where $ (E_{(d)})_{i,j} $ is the $d\times d$ matrix with a $1$ in the $(i,j)$th entry and zeroes everywhere else. So, for a flat connection on the torus with these holonomies, the adjoint fields have gauge fugacities $ \omega_{N/d}^a \; q^\frac{(-1)^a b}{N/d}\frac{z_i}{z_j}$, where $a,b\in \mathfrak{C}_{N/d}$, and $z_i= \exp(2\pi i u_i)$ with $u_i=(\theta_t)_i-\tau (\theta_s)_i$. One of the $d$ modes with $a=b=0$ and $i=j$ corresponds to the identity matrix, and should be excluded, as above for the $d=1$ case. Putting everything together, the contribution from a component of the moduli space isomorphic to $\Mcal_{N,d}$ is schematically
\be \int_{\Mcal_{N,d}} Z_{\rm 1-loop} =d  \frac{1}{(N/d)^2} \frac{1}{d!} \oint_{\mathfrak{M}_{d}}\left( \prod_i d u_i \right)\; \frac{1}{\Xi(0)} \prod_{a,b\in \mathfrak{C}_{N/d}} \prod_{i,j=1}^d \Xi\left(\tfrac{a+(-1)^a b \tau }{N/d} +u_i-u_j\right), \label{eq:integralOnMNd}\ee where the $1/\Xi(0)$ term serves to remove from the product the mode corresponding to the identity element in the Lie algebra.
As a check, we see that this formula reproduces our earlier expression \eqref{eq:McalN1contribution} for $d=1$, and reproduces $1/N$ times the expression for the integral over the $SU(N)$ moduli space obtained by \cite{Gadde:2013dda,Benini:2013nda,Benini:2013xpa} for the integral over the moduli space of the trivial bundle with $d=N$, as can be seen by lifting $\Mcal_d$ to $SU(N)$.


\section{Elliptic genus of $\msym_2$}\label{sec:MSYM elliptic genus}

\subsection{Setup} \label{subsec:MSYM elliptic genus setup}

We are now sufficiently equipped to turn to the computation of the elliptic genus of $\msym_2$. To compute the elliptic genus of a $\Ncal=(8,8)$ supersymmetric theory, it is convenient to pick an $\Ncal=(0,2)$ subalgebra of the $\Ncal=(8,8)$ supersymmetry algebra and express the fields and the Lagrangian in representations of this $\Ncal = (0,2)$ superalgebra. As elaborated in Appendix \ref{apx:action and susy}, a choice of an $\Ncal=(0,2)$ subalgebra is given by picking two right moving supercharges $\Qcal_R^\pm$ that generate right-moving supersymmetry transformations $\varepsilon_R^{\pm}\subset \varepsilon_R^{\alpha}$, such that $\varepsilon_R^\pm$ (and thus $\Qcal_R^\pm$) are eigenstates of a weight of the $\mathbf{8}_s$ representation. To paraphrase the Appendix for convenience, this choice decomposes the R-symmetry group as $Spin(8)\rightarrow Spin(2) \times Spin(6) \cong U(1)_R \times SU(4)$, such that 
\be \begin{split} \mathbf{8}_s &\rightarrow \mathbf{1}_{+1} \oplus \mathbf{6}_0 \oplus \mathbf{1}_{-1}
\\ \mathbf{8}_c &\rightarrow \mathbf{4}_{-\foh} \oplus \bar{\mathbf{4}}_{+\foh}
\\ \mathbf{8}_v &\rightarrow \mathbf{4}_{+\foh} \oplus \bar{\mathbf{4}}_{-\foh}. 
\end{split} \label{eq:so8tosu4split}
\ee Let $\{\pm e^i\}_{i=1,\dots,4} \subset \mathfrak{h}^*$ be the weights of the $\mathbf{8}_v$ representation of $Spin(8)$, and let $\{K_j\}_{j=1,\dots,4}\subset \mathfrak{h}$ denote the Cartan generators with $e^i(K_j) = \delta^i_j$.  A concrete choice of $\varepsilon_R^{\pm}$ is given by the $\mathbf{8}_s$ weights $\pm r$ where $r=\foh ( e_1 + e_2 + e_3 + e_4 )$, for which $U(1)_R$ is generated by the Cartan generator $J_R = \foh(K_1+K_2+K_3+K_4)$. 

Under such a split, the $SU(4)$ factor commutes with the supercharges $\Qcal_R^{\pm}$; therefore it is a flavor symmetry from the perspective of the $\Ncal=(0,2)$ superalgebra. This allows us to define the index in the Ramond-Ramond sector via the $\Ncal=(0,2)$ flavored elliptic genus \cite{Gadde:2013dda,Benini:2013nda,Benini:2013xpa} 
\be
\Tr_{\mathcal{H}} (-1)^{F} q^{H_L} \bar{q}^{H_R} \prod_A a_A^{f_A}
\ee where $f_A$ are the Cartan generators of ${ Spin}(6)\cong SU(4)$. Generalizing the index to include the $\theta$ angle, we obtain
\be
\Tr_{\mathcal{H}} e^{i\theta \int w_2} (-1)^{F} q^{H_L} \bar{q}^{H_R} \prod_A a_A^{f_A} = \sum_k \Tr_{\mathcal{H}_k} e^{i\theta k} (-1)^{F} q^{H_L} \bar{q}^{H_R} \prod_A a_A^{f_A}.
\ee

Under the decomposition \eqref{eq:so8tosu4split}, the fields decompose into $SU(4)$ representations as
\be
\begin{split}
\{X^i\} & \rightarrow \{\phi^A,\bar{\phi}_A\}
\\ \{\chi_L^\alpha\} &\rightarrow \{\lambda_-, \bar{\lambda}_-,\psi_-^{AB}\}
\\ \{\chi_R^{\dot \alpha} \} & \rightarrow \{\psi_+^A, \bar{\psi}_{+A}\},
\end{split} 
\ee which can be reorganized into $\Ncal=(0,2)$ superfields $\{\Phi^A, \bar{\Phi}_A, \Lambda, \bar\Lambda, \Psi^{A4},\bar{\Psi}_{A4}\}$ as
\be
\begin{split}
\Phi^A &= \phi^A + \theta^+ \psi_+^A + \theta^+ \thetabar^+ D_+ \phi^A
\\ \Lambda &= \lambda_- + \theta^+ \frac{1}{\sqrt{2}}(D+iF_{09}) + \theta^+ \thetabar^+ D_+ \lambda_-
\\ \Psi^{A4} &= \psi_+^{A4} +\theta^+ G^{A4} + \thetabar^+ E^{A4}(\Phi) + \theta^+ \thetabar^+ D_+ \psi_+^{A4}.
\end{split} \label{eq:02 superfields}
\ee The Fermi multiplet $\Lambda$ is the $\Ncal=(0,2)$ vector multiplet, and carries the gauge field strength $F_{09}$ (or $\mathcal{F}_{09}$ for the $U(N)+B$ theory).  The $E$-type interaction term is $E^{A4}(\Phi) = - i \sqrt{2} g [\Phi^A, \Phi^4]$.
There is also a $J$-term superpotential
\begin{equation}
ig\Tr \left. \int d\theta^+ \Psi^{A4} J_A(\Phi) \right|_{\thetabar^+=0} + h.c. =  ig\frac{\epsilon_{ABC4}}{3!} \Tr \left. \int d\theta^+ \Psi^{A4} [ \Phi^B, \Phi^C] \right|_{\thetabar^+=0} + h.c. \label{eq:Jterm superpotential}
\end{equation}
Perhaps the easiest way to derive these interactions is from the Lagrangian of 4d $\Ncal = 4$ $\sym$ written in $\Ncal=1$ supermultiplets. When dimensionally reduced to 2d, we get 2d $\Ncal=(8,8)$ $\sym$, expressed in $\Ncal=(2,2)$ vector and chiral superfields, denoted $\tilde \Sigma$ and $\tilde \Phi^{1,2,3}$, respectively, with the $\Ncal=1$ superpotential descending to the $\Ncal=(2,2)$ superpotential
\begin{equation}
ig \Tr \int d\theta^2 \tilde\Phi^1 [\tilde \Phi^2,\tilde \Phi^3] + h.c.. \label{eq:Fterm superpotential}
\end{equation}
Now, we can decompose the $\Ncal=(2,2)$ multiplets and the $\Ncal=(2,2)$ superpotential into their $\Ncal=(0,2)$ counterparts as described in \cite{Gadde:2013sca}. The vector multiplet $\tilde \Sigma$ decomposes into a chiral multiplet $\Phi^4$ and the Fermi vector multiplet $\Lambda$. The chiral multiplet $\tilde \Phi^A$ of $\Ncal=(2,2)$ decomposes into a $\Ncal=(0,2)$ chiral multiplet $\Phi^A$ and Fermi multiplet $\Psi^{A4}$, where the Fermi multiplet has $E$-term $ \Dbar_+ \Psi^A = i\sqrt{2} g [\Phi^4, \Phi^A]$.
The $\Ncal=(2,2)$ superpotential $W(\Phi)$ descends to $J_A(\Phi) = \frac{\partial W}{\partial \Phi^A}$, which reproduces our expression above.
 

For the free $U(1)$ theory, the index as defined vanishes due to the zero mode of $\lambda_-$ and its conjugate, as usual. This is because $\lambda_-$ and its conjugate are in the same eigenstate of bosonic symmetries as the $\Ncal=(0,2)$ supercharges, including the R-symmetry, and have opposite fermion number, so their contributions cancel. 
But, following \cite{Cecotti:1992qh,Gukov:2004fh}, we can remove the contribution from the problematic zero modes by inserting a factor of $J_R$ into the definition of the trace, as we will discuss in detail in Section \ref{sec:symmetric orbifold elliptic genus}. Then the index is simply the product of the one loop partition functions for each of the superfields
\ba
\mathcal{I}_{U(1)} &= Z_\Lambda  \prod_A Z_{\Phi^A}  Z_{\Psi^{A4}} 
= \eta(\tau)^3  \frac{\prod_{A=1}^3 \theta_1(\tau | \xi_A+\xi_4) }{\prod_{A=1}^4 \theta_1 (\tau | \xi_A)} \label{eq:U1 index}
\ea where $\xi_A$ are holonomies for flat background gauge fields for the $ SU(4)$  \lq\lq{}flavor\rq{}\rq{} symmetry, coupling to fields via $\rho(\xi) = \rho^A \xi_A$, where $\rho$ is a weight of the fundamental $SU(4)$ representation. The holonomies $\xi_A$ satisfy
\begin{equation}
\sum_A \xi_A = 0,
\end{equation} which is the determinant constraint of $SU(4)$, or equivalently the superpotential constraint. The Dedekind eta function is defined as 
\be \eta(\tau) = q^{1/24} \prod_{n=1}^\infty (1-q^n), \ee and the Jacobi theta function is defined as
\be \theta_1(\tau|u) = -i q^{1/8}z^{1/2} \prod_{n=1}^\infty (1-q^n)(1-z q^n)(1-z^{-1} q^{n-1}), \ee with $q=e^{2\pi i \tau}$ and $z=e^{2\pi i u}$.

Let\rq{}s recall that in order to compute the index for the interacting gauge theory, one also needs to introduce gauge fugacities, and then impose Gauss\rq{} Law, which takes the form of a contour integral. Since the theory is free in the UV, and the index is scale invariant, we can do the computation in the free UV limit, so we only need the contribution from each free field. The integrand of the contour integral for the gauge theory index is then 
\be Z_{\rm 1-loop}(\tau|u;\xi) = \prod_\alpha \Xi(\tau|\alpha(u); \xi), \ee
where $\Xi(\tau | \alpha(u);\xi)$ is the factor from the modes with charge $\alpha$ in the presence of a background flat gauge connection specified by $u$, with $\alpha(u)$ as discussed in Section \ref{subsec:charge of adjoint fields} for the various components of the moduli space of flat connections. For $\msym_2$, the free field index is
\be \Xi(\tau | u; \xi) := \frac{ \theta_1(\tau | u) \prod_{A=1}^3 \theta_1(\tau | \xi_A + \xi_4 + u)}{\prod_{A=1}^4 \theta_1 (\tau | \xi_A + u)}. \ee  
Note that we can recover the $U(1)$ index as 
\be \mathcal{I}_{U(1)}(\tau|\xi) = \left. -\frac{\partial}{\partial u}\right|_{u=0} \Xi(\tau | u; \xi). \ee

The function $\Xi(\tau | u; \xi)$ inherits the following periodicity properties from the theta function $\theta_1(\tau|u)$,
\begin{equation} 
\begin{split} \Xi(\tau | u + a +b\tau ; \xi) &=e^{-2\pi i b (2 \xi_4)} \Xi(\tau | u; \xi),
\\ \Xi(\tau | u; \xi_1 + a+b\tau,\xi_2,\xi_3) &= e^{2\pi i b (2u)}\Xi(\tau | u; \xi),
\end{split} \end{equation}
as well as the following modular transformation properties,
\begin{equation} 
\begin{split} \Xi(\tau+1 |u ; \xi) &=\Xi(\tau |u ; \xi),
\\ \Xi\left(\left. -\frac{1}{\tau} \right| \frac{u}{\tau}; \frac{\xi}{\tau}\right) & = e^{\frac{\pi i}{\tau} (4u\xi_4)} \Xi(\tau | u; \xi).
\end{split} \end{equation} These properties imply that the integrand $Z_{\rm 1-loop}(\tau|u;\xi)$, and therefore the index is a modular invariant symmetric Abelian (multi-periodic) function of the variables $\xi_{1,2,3}$ with modular parameter $\tau$. We will explore such functions in Section \ref{sec:symmetric orbifold elliptic genus}, and their uniqueness properties will help us match the gauge theory index to the symmetric orbifold index in Section \ref{subsec:JKRes on trivial bundle}.


\subsection{Contribution from isolated flat connections}

We are now ready to compute the various contributions to the $SU(N)/\Z_N$ gauge theory index from the components of $\Mcal_{\rm flat}$. Let\rq{}s start with the pointlike components, corresponding to isolated  flat connections of the bundles $P_{N,k}$ with $k\perp N$. Applying our earlier result  \eqref{eq:McalN1contribution}, we have
\be \left. Z_{\rm 1-loop} \right|_{\Mcal_{N,1}} = \frac{1}{N^2} \prod_{\substack{a,b\in\mathfrak{C}_N \\ (a,b)\ne(0,0)}} \Xi(\tau | \tfrac{a+(-1)^a b\tau}{N}; \xi_A).  \ee
In fact, this expression simplifies quite a bit, due to the identity 
\be  \prod_{\substack{a,b\in\mathfrak{C}_N}} \Xi(\tau | u+ \tfrac{a+(-1)^a b\tau}{N}; \xi_A) =  \Xi(\tau | N u ; N\xi_A). \label{eq:identityMN1Simplification}\ee
We can now rewrite the contribution to the index as
\be \left. Z_{\rm 1-loop} \right|_{\Mcal_{N,1}} = \frac{1}{N^2} \lim_{u\rightarrow 0} \frac{ \Xi(\tau | N u ; N\xi_A)}{ \Xi(\tau | u ; \xi_A)} = \frac{1}{N}\frac{\mathcal{I}_{U(1)}(\tau|N\xi_A)}{\mathcal{I}_{U(1)}(\tau|\xi_A)}.\ee

\subsection{Integral over flat connections on the trivial bundle} \label{subsec:JKRes on trivial bundle}

Let\rq{}s move on to the contributions from components of $\Mcal_{\rm flat}$ of positive dimension. We will start with the component corresponding to the trivial $SU(N)/\Z_N$-bundle $P_{N,0}$, which will be the bulk of our computation. As discussed in Section \ref{subsec:integral over mflat}, we can lift the integral on the moduli space of flat connections $\Mcal_{N,N}$ of $P_{N,0}$ to an integral on the moduli space of flat $SU(N)$-connections $\tilde{\mathfrak{M}}/\mathbb{S}_N$. This allows us to use the formula obtained by \cite{Benini:2013xpa} (see also \cite{Gadde:2013dda,Benini:2013nda}) and write the integral in \eqref{eq:trivial bundle integral lift to SU(N)} as
\be \oint_{\Mcal_{N,N} } Z_{{\rm 1-loop}}(u) = \frac{1}{|\pi_1(SU(N)/\Z_N)|} \frac{1}{|\mathbb{S}_N|} \sum_{u_* \in \tilde{\mathfrak{M}}_{{\rm sing}^*} } \JKRes_{u=u_*}(\mathsf{Q}(u_*),\eta)Z_{{\rm 1-loop}}(u), \label{eq:JKResMNN}\ee where 
\be 
Z_{\rm 1-loop} =  \left(\mathcal{I}_{U(1)}\right)^{N-1} \prod_{i\ne j}\frac{ \theta_1(\tau | u_i-u_j)\prod_{A=1}^3 \theta_1(\tau|\xi_A + \xi_4+u_i-u_j)}{\prod_{A=1}^4 \theta_1 (\tau|\xi_A + u_i - u_j)} \bigwedge_{i=2}^{N} du_i. \label{eq:JKRes integrand}
\ee 

The authors of \cite{Benini:2013xpa} give a detailed prescription for evaluating the $\JKRes$ operation. Here, we will briefly recall parts of the prescription, and compute the residue. Let $r$ denote the rank of the gauge group, so $r=N-1$ here for $SU(N)$. The integrand $Z_{\rm 1-loop}$ is naturally a meromorphic $(r,0)$-form on $\tilde{\mathfrak{M}}$, which is the torus $\mathfrak{h}_\CC /(Q^{\vee} + \tau Q^{\vee}) \cong (\CC/\Z+\tau\Z)^{r}$, where $\mathfrak{h}$ is the Cartan subalgebra of $SU(N)$, and $Q^\vee$ is the coroot lattice. We pick $u_2,\dots,u_{N}$ as coordinates on $\tilde{\mathfrak{M}}$ and solve for $u_1$ using the trace constraint $\sum_i u_i =0$. We observe that $Z_{\rm 1-loop}$ is singular along the hyperplanes 
\be H_{ij}^{A} = \{u_i-u_j + \xi_A = 0 \mod \Z+\tau \Z \} \subset \tilde{\mathfrak{M}}. \ee 
 Let $Q_{ij}^{A}\in \mathfrak{h}^*$ denote the weight of the multiplet responsible for the hyperplane $H_{ij}^{A}$, which are the non-zero roots $Q_{ij}^{A}(u) = u_i-u_j$. Let $\mathsf{Q}(u_*) =\{Q_{ij}^{A} \; |\; u_* \in H_{ij}^{A} \} $ denote the set of charges of the singular hyperplanes meeting at $u_*$. The collection of points $u_*$ where at least $r$ singular hyperplanes intersect is denoted by $\tilde{\mathfrak{M}}_{\rm sing^*}$. When the charges $\mathsf{Q}(u_*)$ of all singular hyperplanes meeting at a point are contained in a half-space of $\mathfrak{h}^*$, the arrangement of hyperplanes is termed \lq\lq{}projective\rq\rq{}. When there are exactly $r$ singular hyperplanes intersecting at a point, labeled say $H_{j_1},\dots,H_{j_r}$, the arrangement is termed ``non-degenerate''. To evaluate the residue, we need to pick a covector $\eta \in \mathfrak{h}^*$, which for theories with only adjoint fields specifies a Weyl chamber. For a projective and non-degenerate arrangement, the residue is determined by the operation
\be \JKRes_{u=u_*}(\mathsf{Q}(u_*),\eta) \frac{du_1\wedge \cdots \wedge du_r}{Q_{j_1}(u-u_*)\cdots Q_{j_r}(u-u_*)} = \begin{cases} {\large \frac{1}{|\det(Q_{j_1}\dots Q_{j_r})|} } & \mbox{if } \eta\in {\rm Cone}(Q_{j_1}\dots Q_{j_r}), \\ 0 &\mbox{otherwise.} \end{cases} \label{eq:defJKRes} \ee 
Here, ${\rm Cone}(Q_{j_1}\dots Q_{j_r})$ stands for the positive cone generated by the charge rays $Q_{j_1},\dots, Q_{j_r}$.
When the arrangement is degenerate, so there are more than $r$ singular hyperplanes intersecting, the $\JKRes$ operation is more complicated, as one needs to specify the precise cycle to integrate on. However, for the case of interest for us, whenever the arrangement is degenerate, one can exploit the linearity of the $\JKRes$ operation to determine the cycle relatively easily, as was pointed out in some examples in \cite{Benini:2013xpa}. In any case, the $\JKRes$ operation corresponds to a particular linear combination of iterated residues, and in our case we will be able express $\JKRes$ explicitly as a somewhat simple prescription of iterated residues.

Let\rq{}s analyze which poles give non-zero contributions to the sum in \eqref{eq:JKResMNN}. 
It simplifies the classification of poles to note that non-zero residues are from points $u_{*}$ where $s$ singular hyperplanes and $s'$ zero hyperplanes intersect, such that $s-s' = r$.
We see that $Z_{\rm 1-loop}$ has zeroes along the hyperplanes defined by
\be \begin{split} N_{ij} &= \{u_i - u_j = 0 \mod \Z+\tau \Z \}, 
\\ N_{ij}^{B4} &=\{u_i - u_j + \xi_B + \xi_4=0 \mod \Z+\tau \Z\}, \end{split} \ee
for $i\ne j$ and $B=1,2,3$. So, for example, at the $N^2$ points where the hyperplanes $H_{i+1,i}^{A}$ with $i=1,\dots,N-1$ and some fixed $A$ intersect, there are no other singular or zero hyperplanes intersecting (for generic $\xi_A$). These points therefore give non-zero contributions as long as $\eta\in {\rm Cone}(\{Q_{i+1,i}^{A}\}_{i=1,\dots,N-1})$. However, whenever say $H_{i,j}^{A}$ and $H_{i\rq{}, j}^{A}$ intersect, we have $u_i=u_{i\rq{}}$, at which point there is a double zero in the integrand, and such points don\rq{}t contribute for generic $\xi_A$. 

We note that sets of hyperplanes that contribute a non-zero residue always intersect at $N^2$ points, and each of these points will contribute identical residues. 
This is coming from the fact that we have lifted the integral on the trivial $SU(N)/\Z_N$-bundle\rq{}s moduli space to the $SU(N)$ moduli space $\tilde{\mathfrak{M}}$, which as we discussed in Section \ref{subsec:integral over mflat} is an $N^2\rightarrow1$ covering. For concreteness, we will continue the integral on $\tilde{\mathfrak{M}}$ to make direct contact with the literature, and observe that we will obtain $N^2$ times the integral over the $SU(N)/\Z_N$ moduli space.

Let\rq{}s return to the classification of poles. 
There are some points where a degenerate intersection occurs with the required number of zero hyperplanes for the residue to be non-zero. When this is the case, first of all, we need to determine what order of iterated residues $\JKRes$ corresponds to. A second point that needs attention is as follows. We note that due to the constraint $\sum \xi_A = 0$, the second set of zero hyperplanes $N_{ij}^{B4}$ can be written as  \be N_{ij}^{AB} = \{u_i - u_j + \xi_A + \xi_B=0 \mod \Z+\tau \Z\} \ee with $A,B=1,2,3,4$, but $A\ne B$ --- essentially, as an rank 2 antisymmetric tensor of $SU(4)$. Although the zeroes are totally symmetric in the $\xi_A$ (as expected, since the integrand is totally symmetric in the $\xi_A$), the signs of the factor in the integrand giving these hyperplanes differ for the pairs $(A,B) \in  \{(1,4),(2,4),(3,4)\}$ versus $(A,B) \in  \{(1,2),(1,3),(2,3)\}$. This introduces a subtle sign in the computation of the residue, which we have to keep track of.

For concreteness, let\rq{}s look closely at an example, as it will illuminate some of the subtleties in the computation. For $N=4$, there are $N^2=16$ points where four singular hyperplanes $H_{12}^{A}$, $H_{13}^{B}$, $H_{24}^{B}$, and  $H_{34}^{A}$ meet the zero hyperplane $\{\epsilon(A,B) (u_1-u_4)+\xi_A+\xi_B=0\}$. Here, $\epsilon(A,B)$ is the sign that determines the correct zero hyperplane, $N_{14}^{AB}$ or $N_{14}^{BA}$; it is $1$ if either of $A$ or $B$ is 4, and $-1$ otherwise. %
The intersection occurs at the points \begin{equation} (u_2,u_3,u_4)=\frac{1}{2}(\xi_A-\xi_B,-\xi_A+\xi_B,\xi_A+\xi_B) + \frac{a+b\tau}{4}(1,1,1) \end{equation} for $a,b=1,\dots,4$. A more suitable choice of coordinates is given by $v_i =Q_{i1}(u)= u_{i} - u_1$ for $i=2,3, 4$. The intersection points in these coordinates are at \begin{equation} (v_2,v_3,v_4)=(\xi_A,\xi_B,\xi_A+\xi_B) + (a+b\tau)(1,1,1). \end{equation} First of all, let\rq{}s note that the integrand is doubly periodic in each of the variables $v_i$ under translations by $\Z+\tau\Z$, so each of the poles contributes the same residue. Shifting the coordinates so that the intersection happens at $v_i=0$, we need to evaluate
\begin{equation} \JKRes_{v=0}(\mathsf{Q}_*,\eta) \frac{\epsilon(A,B) v_4}{v_2 v_3 (v_4-v_2) (v_4-v_3)} \frac{dv_2\wedge dv_3\wedge dv_4}{4}. \label{eq:JKResSquarePole}
\end{equation}
The set of charges $\mathsf{Q}_*$ is $\{Q_{12},Q_{13},Q_{24},Q_{34}\}$, which are \begin{equation} Q_{12}= (-1,0,0),\quad Q_{13}=(0,-1,0),\quad Q_{24} = (1,0,-1),\quad Q_{34} =(0,1,-1) \end{equation} in coordinates dual to $v_i$.
We pick the convenient choice of $\eta = (-1,-1,-1)$ in these coordinates. Now, we need to determine which cycle of integration $\JKRes$ corresponds to for this $\eta$. As discussed in \eqref{eq:JKResMNN}, there could be various such cycles, depending on which sub-chamber $\eta$ sits in; however, the results are equivalent. By linearity of the $\JKRes$ operation, if we find some cycle of integration such that when applied to the 3-form defined by
\begin{equation} \omega_{234} =\left(\frac{a}{v_2v_3(v_4-v_2)}+\frac{b}{v_2v_3(v_4-v_3)}+\frac{c}{v_2(v_4-v_2)(v_4-v_3)}+\frac{d}{v_3(v_4-v_2)(v_4-v_3)}\right)
\end{equation} gives the correct residue for each of the linear pieces, as according to \eqref{eq:defJKRes}, then it is the right prescription for the degenerate case. Noting that for the four subsets of charges, only ${\rm Cone(Q_{12},Q_{13},Q_{24})}$ and ${\rm Cone(Q_{12},Q_{13},Q_{34})}$ contain $\eta$, the correct cycles are determined as $\Res_{v_4=0} \Res_{v_3=0} \Res_{v_2=0}$ and $\Res_{v_4=0} \Res_{v_2=0} \Res_{v_3=0}$, as both evaluate to 
$ a+b$ when applied to $\omega_{234}$. Therefore, applying either of the iterated residues to \eqref{eq:JKResSquarePole}, we see that it evaluates to $\epsilon(A,B)/4$. Such poles generalize to $N>4$ as Young tableaux along pairs $(A,B)$ as one expects. 

Another subtlety comes from poles containing ``cubes'', which starts occurring for $N\ge8$. Concretely, for $N=8$, we have a pole at the point
\begin{equation} (v_i)_{i=2,\dots,8} = (\xi_1,\xi_2,\xi_3,\xi_1+\xi_2, \xi_1+\xi_3,\xi_2+\xi_3, \xi_1+\xi_2+\xi_3).\end{equation} There are $13$ singular hyperplanes 
\be H_{12}^{1}, H_{13}^{2}, H_{14}^{3}, H_{25}^{1}, H_{35}^{2}, H_{26}^{3}, H_{46}^{1}, H_{37}^{3}, H_{47}^{2}, H_{58}^{3}, H_{68}^{2}, H_{78}^{1}, H_{81}^{4} \ee
and $6$ zero hyperplanes $N_{51}^{34}, N_{61}^{24}, N_{71}^{14}, N_{82}^{14}, N_{83}^{24}, N_{84}^{34} $
meeting at this point. However, the charge vector $Q_{81}$ coming from $H_{81}^{4}$ points outside of any half-space containing all the other charge vectors, so the arrangement is not projective. As was pointed out in \cite{Benini:2013xpa}, we can deal with this situation by relaxing the constraint on the R-symmetry fugacities (which resolves the intersection into a bunch of projective ones), computing the residues, and then taking the limit $\epsilon\rightarrow 0$. Relaxing the constraint on $\xi_A$ to $\xi_1+\xi_2+\xi_3+\xi_4 = \epsilon$, the singular point is resolved to two points, at $v_8 = \xi_1+\xi_2+\xi_3$ and at $v_8 = -\xi_4 = -\xi_1-\xi_2-\xi_3 + \epsilon$ with $v_2,\dots,v_7$ as before. For $\eta = (-1,\dots, -1)$, the second point does not contribute, and to obtain the contribution from the first point, we need to calculate
\begin{equation}
\begin{split} \JKRes_{v=0} (\mathsf{Q}_*,\eta) &\frac{(v_5+\epsilon)(v_6+\epsilon)(v_7+\epsilon)}{v_2v_3v_4(v_5-v_2)(v_5-v_3)(v_6-v_2)(v_6-v_4)(v_7-v_3)(v_7-v_4)} 
\\ &\qquad \times \frac{(v_8-v_2+\epsilon)(v_8-v_3+\epsilon)(v_8-v_4+\epsilon)}{(v_8-v_5)(v_8-v_6)(v_8-v_7)(-\epsilon-v_8)} \frac{\bigwedge_{i=2}^{8} dv_i}{8}. \label{eq:nonprojective residue for cube}
\end{split}
\end{equation} We can determine possible choices of a cycle of integration for this degenerate arrangement as above, and once again the residue is independent of this choice. One choice is given by
\begin{equation}
\JKRes_{v=0} (\mathsf{Q}_*,\eta) \bigwedge_{i=2}^{8} dv_i = \Res_{v_8=0} \Res_{v_7=0} \dots \Res_{v_2=0},
\end{equation} so \eqref{eq:nonprojective residue for cube} evaluates to $-1/8$. Note that this sign comes from the singular hyperplane $H_{18}^{4}$ with the problematic charge covector which made the arrangement non-projective in the first place, and is separate from the sign coming from zero-hyperplanes discussed above. So, in general we need to keep track of both sources of sign for the residue.



Finally, we note that starting $N\ge 16$, there are poles containing ``hypercubes'', with $v_{16} =\xi_1+\xi_2 +\xi_3 +\xi_4$. Due to the constraint on $\xi_A$, $v_{16}= 0$ and there is a double zero from $N_{16,1}$ and $N_{1,16}$, so such poles have vanishing residue.

We are now ready to compute the contour integral for general $N$.
The contributing poles in any Weyl chamber are classified by certain 4d Young tableaux of size $N$.\footnote{4d Young tableaux of size $N$ also classify solid (3d) partitions of $N$, $ \sum_{i,j,k} n_{i,j,k} = N, $ where for each nonzero $n_{ijk}$, there are $n_{ijk}$ corresponding nodes $(i-1,j-1,k-1,l)$, with $0\le l < n_{ijk}$. In \cite{atkin1967some}, such partitions are denoted 4d partitions of $N$.}
A 4d Young tableau 
is a collection of $N$ ``nodes'' $Y=(y_1,\dots, y_N) \in \Z^4_{\ge 0}$, subject to the ``stacking'' condition: if the node $x=(x^1,x^2,x^3,x^4)\in Y$, then so do all the nodes $y=(y^1,y^2,y^3,y^4)$ with $0\le y^A \le x^A$ for all $A=1,2,3,4$ \cite{atkin1967some}. We also require that each node $y_i$ have at most 3 non-zero coordinates $y_i^A$.
We will denote the collection of such 4d Young tableaux of size $N$ by $\mathcal{Y}_N$. Each such 4d Young tableau $Y$ of size $N$ describes $N! \cdot N^2$ poles of the integrand, at coordinates given by solutions to $u_i-u_j = y_{\sigma(i)}^A \xi_A, $ for some choice of $j$ and the $(N-1)!$ orderings $\sigma(i)$ of the remaining $u_i$ with $i\ne j$.\footnote{We have picked $y_j^A = 0$ which we are free to do for any $Y$.} The choice of $j$ is related to the choice of a Weyl chamber; for any choice of $\eta$ only $(N-1)! \cdot N^2$ poles survive the $\JKRes$ operation, corresponding to some fixed $j$. For concreteness, we fix $j=1$ with the convenient choice of $\eta= (-1,-1,\dots,-1)$ in coordinates $(u_2,u_3,\dots,u_N)$. Since the integrand is symmetric in the $u_i$, the $(N-1)!$ orderings $\sigma(i)$ contribute identically, cancelling part of the factor coming from the order of the Weyl group. We define $v_i =Q_{i1}(u)= u_{i} - u_1$ for $i=2,\dots, N$, noting the relation $\sum u_i = 0$. Contributing poles are at points $v(Y)$ given by coordinates $v_i=y_i^A \xi_A + {a+b\tau}$, for $a,b=1,\dots,N$. Due to the periodicity structure of the integrand, the sum over $a,b$ is trivial and produces a factor of $N^2$.

We introduce the following partial ordering $\preceq$ on the nodes of 4d Young tableaux,
\begin{equation}
y_i\preceq y_j \quad \mbox{ if } y_i^A \le y_j^A \mbox{ for all } A,
\end{equation} which keeps track of the stacking of the nodes.
The operation $\JKRes$ for a pole $Y=(y_1,\dots,y_N)$, partially ordered such that $y_i\preceq y_j$ if $i<j$, is given explicitly by the iterated residue
\begin{equation}
\JKRes_{u=u_*}(\mathsf{Q}_*,\eta) {\bigwedge du_i} = \frac{1}{N}\Res_{v_N = y_N^A\xi_A} \cdots \Res_{v_{3} = y_{3}^A\xi_A}\Res_{v_2 = y_2^A\xi_A}.
\end{equation}

The integral over the moduli space is then
\ba 
\oint_{\Mcal_{N,N}} Z_{\rm 1-loop} &= \frac{1}{N} \sum_{Y \in \mathcal{Y}_N} N^2 \JKRes_{v_i= y_i^A \xi_A} (\mathsf{Q}_*,\eta) \; Z_{\rm 1-loop} (u) 
\\ &= \frac{1}{N} \sum_{Y \in \mathcal{Y}_N} \epsilon(Y) \lim_{\delta \rightarrow 0} \frac{1}{\Xi(\tau|\delta;\xi)}\prod_{i,j} \Xi(\tau | y_i^A \xi_A-y_j^A \xi_A + \delta;\xi),
\ea
where we have introduced an auxiliary variable $\delta$ to simplify the expressions of the residues. The coefficient $\epsilon(Y)$ is a sign due to degenerate and non-projective intersections, and is determined as follows. Let $c_3(Y)$ be the number of nodes in $Y$ with at least 2 nonzero entries in the first 3 coordinates, and let $c_4(Y)$ be the number of nodes in $Y$ with exactly 3 nonzero coordinates, or
\begin{equation}
\begin{split} c_3(Y)&= \#\{y_i \in Y\ | \ y_i^B = 0 \mbox{ for at most one $B$, with } B\in\{1,2,3\}.\}
\\ c_4(Y) &= \#\{y_i \in Y\ | \ y_i^A = 0 \mbox{ for exactly one $A$, with } A\in\{1,2,3,4\}.\}
\end{split}
\end{equation} Then, the sign $\epsilon (Y)$ is given by
\begin{equation}
\epsilon(Y) = (-1)^{c_3(Y) + c_4(Y)}.
\end{equation}

We conjecture that the sum over the residues greatly simplifies to the expression
\be\frac{1}{N} \sum_{|Y|=N} \epsilon(Y) \lim_{\delta \rightarrow 0}\frac{1}{\Xi(\tau|\delta;\xi)}\prod_{i,j} \Xi(\tau | y_i^A \xi_A-y_j^A \xi_A + \delta;\xi) =\frac{1}{N} \sum_{s|N} s \frac{\mathcal{I}_{U(1)}(\tau|\tfrac{N}{s}\xi)}{\mathcal{I}_{U(1)}(\tau|\xi)}. \label{eq:index simplification} \ee This is a highly nontrivial simplification to check analytically, as the summands on the left-hand side grow in number and complexity very quickly in $N$. Fortunately, the functions on both sides of this equation are very special, and they enjoy some very restrictive properties, which allows us to make some exact statements. Specifically, they are modular invariant symmetric Abelian (multi-periodic) functions of the variables $\xi_{1,2,3}$ with the modulus $\tau$ and period as in \eqref{eq:period matrix}, of the kind explored in detail in Section \ref{sec:symmetric orbifold elliptic genus}. This follows from the periodicity and modular transformation properties of $\Xi(\tau|u;\xi)$ and $\mathcal{I}_{U(1)}(\tau|\xi)$; as the integrand \eqref{eq:JKRes integrand} is such a function, so is the integral. We will explore some key properties of such functions in Section \ref{sec:symmetric orbifold elliptic genus}, leading up to Lemma \ref{lemma:ModularInvariantAbelianFunction} which states that such functions are completely determined by the rational function in variables $a_{1,2,3} = \exp 2\pi i \xi_{1,2,3}$ obtained by setting $\tau = i\infty$ (or $q=0$), corresponding to the constant term in the Fourier expansion in $q$. This dramatically simplifies the effort of checking \eqref{eq:index simplification}, since if we can show the equality for $q=0$, the full equality follows exactly! We were able to show this for $N\le 7$ by using \emph{Mathematica} to simplify the sum over the residues with $q=0$.  For larger $N$, up to $N\le12$, we checked that the pole structure of the rational functions obtained by setting $q=0$ on both sides agrees, as well as by performing some numerical checks. 


\subsection{Integral over flat connections on generic bundles} 
Having computed the integral on the moduli space of the trivial bundle, turns out we can infer the integral on each of the other components of $\Mcal_{\rm flat}$. 
We first note that we can use the identity \eqref{eq:identityMN1Simplification} to simplify the integrand in \eqref{eq:integralOnMNd},
\begin{equation} \int_{\Mcal_{\mathrlap{N,d}}}  Z_{\rm 1-loop} =\frac{d^2}{N} \frac{\mathcal{I}_{U(1)}(\tau|\tfrac{N}{d}\xi)}{\mathcal{I}_{U(1)}(\tau|\xi)} \frac{1}{d!} \oint_{\mathfrak{M}_{d}}\left( \prod_i d u_i \right)\left(\frac{N}{d}\mathcal I_{U(1)} \left(\tau |\tfrac{N}{d}\xi \right)\right)^{d-1} \prod_{\substack{i,j=1\\i\ne j}}^d \Xi(\tau|{\tfrac{N}{d}}(u_i-u_j);\tfrac{N}{d}\xi). \end{equation} We recognize the first factor as the contribution from $\Mcal_{{N}/{d},1}$. The integral is the same as the integral over $\Mcal_{d,d}$, but with scaled flavor charges $\xi \rightarrow \frac{N}{d} \xi$. Quoting our result above, we have 
\be \int_{\Mcal_{N,d}} Z_{\rm 1-loop} = \frac{1}{N} \frac{\mathcal{I}_{U(1)}(\tau|\tfrac{N}{d}\xi)}{\mathcal{I}_{U(1)}(\tau|\xi)} \sum_{s|d} s \frac{\mathcal{I}_{U(1)}(\tau|\tfrac{d}{s} \tfrac{N}{d}\xi)}{\mathcal{I}_{U(1)}(\tau|\tfrac{N}{d}\xi)} = \frac{1}{N}  \sum_{s|d} s \frac{\mathcal{I}_{U(1)}(\tau|\tfrac{N}{s}\xi)}{\mathcal{I}_{U(1)}(\tau|\xi)}. \ee

\subsection{Putting the pieces together}
Adding up the contributions from each of the components of the moduli space of flat connections, we obtain the index
\be \mathcal{I}_{SU(N)/\Z_N}^\theta(\tau|\xi) = \frac{1}{N}\sum_{k=1}^N e^{i \theta k} \sum_{s|\gcd(k,N) } s \frac{\mathcal{I}_{U(1)}(\tau|\tfrac{N}{s}\xi)}{\mathcal{I}_{U(1)}(\tau|\xi)}. \ee In fact, we can evaluate the sum over $k$ with given  $\theta  = \frac{2\pi M}{N}\pmod{2\pi}$
\be \mathcal{I}_{SU(N)/\Z_N}^{\theta = \frac{2\pi M}{N}} (\tau|\xi)= \sum_{s|D} \frac{\mathcal{I}_{U(1)}(\tau|{s}\xi)}{\mathcal{I}_{U(1)}(\tau|\xi)} = \frac{\mathcal{I}_{D}}{\mathcal{I}_1}(\tau|\xi) \ee where $D=\gcd (M, N)$. Thus we establish that the index for the $SU(N)/\Z_N$ $\msym_2$ theory at theta angle $\theta  = \frac{2\pi M}{N}$ is equal to the index of the sigma model into $(\R^8)^{D-1}/\mathbb{S}_D$, providing strong evidence that the IR limit of the gauge theory with the corresponding theta parameter is described by this sigma model.

We can also easily infer the index of the $SU(N)$ and $SU(N)/\Z_K$ theories for each $K|N$ with our results thus far. For each such theory, the contributing bundles are a subset of the $SU(N)/\Z_N$-bundles, with the moduli space of flat connections lifted appropriately. For the $SU(N)$ theory only the trivial bundle contributes, so we have the index
\be \mathcal{I}_{SU(N)} = \sum_{s|N} s\frac{\mathcal{I}_{U(1)}(\tau|\tfrac{N}{s}\xi)}{\mathcal{I}_{U(1)}(\tau|\xi)} = \sum_{k=1}^N \frac{\mathcal{I}_{\gcd(k,N)} }{\mathcal{I}_1} (\tau|\xi),\ee which is the sum of the index of each of the $N$ superselection sectors in the theory, with the $k$th superselection sector described by the sigma model into $(\R^8)^{d-1}/\mathbb{S}_d$ with $d=\gcd(k,N)$. For a $SU(N)/\Z_K$ theory, there are $K$ bundles to sum over, corresponding to those $SU(N)/\Z_N$ bundles with $w_2$ liftable to $H^2(T^2,\Z_K)$ where $\Z_K \subset \Z_N$ --- essentially those with $K|w_2$. Accounting for the volume of gauge transformations and adding in the $\Z_K$-valued $\theta$ angle $\theta = 2 \pi M/N$ with $M\in \Z_{N}/\Z_{N/K}\cong \Z_K$, we obtain the index
\begin{align}
\mathcal{I}_{SU(N)/\Z_K}^{\theta = \frac{2\pi M}{N}} (\tau|\xi)&= \frac{1}{K}\sum_{k=1}^{N/K} e^{i \theta k K} \sum_{s|\gcd(k K,N) } s \frac{\mathcal{I}_{U(1)}(\tau|\tfrac{N}{s}\xi)}{\mathcal{I}_{U(1)}(\tau|\xi)} 
\\ &= \sum_{k\equiv M \! \! \pmod K} \sum_{s|\gcd(k,N)} \frac{\mathcal{I}_{U(1)}(\tau|{s}\xi)}{\mathcal{I}_{U(1)}(\tau|\xi)} 
\\ &= \sum_{k \equiv M \! \! \pmod K} \frac{\mathcal{I}_{\gcd(k,N)}}{\mathcal{I}_1}(\tau|\xi) .
\end{align} For each of the $K$ values of $\theta$, the index is the sum of the indices of the $N/K$ superselection sectors of the $SU(N)$ theory with the same $\Z_{K}$ charge.

As discussed, the index of the $U(N)$ theory can be inferred from that of the $SU(N)$ theory, and is
\begin{equation} \mathcal{I}_{U(N)} (\tau|\xi) = \mathcal{I}_{U(1)} \mathcal{I}_{SU(N)} (\tau|\xi) = \sum_{k=1}^{N} \mathcal{I}_{\gcd(k,N)}(\tau|\xi). \end{equation}
The $U(N)$ theory has $N$ superselection sectors, as expected. 

The index of the $N$ D1-branes worldvolume theory with $U(N)$ gauge field and the $B$-field in the sector with $M$ units of flux $\tilde{c}_1$ is
\begin{equation} \mathcal{I}_{U(N)+B}^M(\tau|\xi) = \mathcal{I}_{U(1)}^{\tilde c_1 = M} \mathcal{I}_{SU(N)}^{\theta = \frac{2\pi M}{N}} (\tau|\xi) = \mathcal{I}_{D}(\tau|\xi). \end{equation}
We have used the fact that the $U(1)$ factor is free, and since the field strength does not contribute to the index, $\mathcal{I}_{U(1)}^{\tilde{c}_1} = \mathcal{I}_1$ in the appropriate topological sector of the supersymmetry algebra. The index summing over all flux sectors (and therefore all BPS sectors) is 
\begin{equation}
\mathcal{I}_{U(N)+B}(\tau|\xi) = \sum_{M\in \Z} e^{iM\phi} \mathcal{I}_D (\tau|\xi).
\end{equation}
Once again, we note that the D1-brane index is invariant under the S-duality of the Type IIB string, which is generated by exchanging $M$ and $N$ and shifting $M$ by a multiple of $N$, while leaving $D$ invariant.

\section{Elliptic genera of $\Ncal=(8,8)$ sigma models
} \label{sec:symmetric orbifold elliptic genus}

We have thus far computed an $\Ncal=(8,8)$ analog of the elliptic genus of the $SU(N)$ and the $U(N)$ $\msym_2$, and claimed that they are equal to the corresponding elliptic genus of some symmetric orbifolds of the supersymmetric sigma model into $\R^8$. In this section, we will compute the elliptic genus of the orbifold sigma model, and establish some of its key properties that allow us to match it with the gauge theory elliptic genus.

\subsection{Elliptic genus of the $\R^8$ sigma model}
For brevity, we will denote the supersymmetric sigma model into $\R^8$ by $\mathcal{C}$. $\mathcal{C}$ is a free theory. When viewed as a non-supersymmetric theory, $\mathcal{C}$ carries 3 $Spin(8)$ flavor symmetry groups, labeled $K_b$, $K_l$, and $K_r$, each acting separately on the 8 real bosons, the 8 real left-moving fermions, and the 8 real right-moving fermions. When viewed as a $\Ncal=(8,8)$ supersymmetric theory, these actions are combined into a single copy of $Spin(8)$, $K$, which is the R-symmetry identified as the rotation symmetry of the target space, with the bosons, the left-moving fermions, and the right-moving fermions transforming in the $\mathbf{8}_v$, $\mathbf{8}_s$, and $\mathbf{8}_c$ representations, respectively (up to $Spin(8)$ triality). We can pick the representations of the fields under $K_b \times K_l \times K_r$ as
\be (\mathbf{8}_v,\mathbf{1},\mathbf{1})\oplus (\mathbf{1},\mathbf{8}_s,\mathbf{1})\oplus (\mathbf{1},\mathbf{1},\mathbf{8}_c). \ee With this choice, $K$ is identified as the diagonal combination of $K_b \times K_l \times K_r$.

The philosophy for computing the flavored elliptic genus is to pick an $\Ncal =(0,2)$ supersymmetry, and insert into the trace fugacities for every bosonic charge which commutes with the chosen supersymmetry. We can think of $\mathcal{C}$ as an $\Ncal=(0,8)$ theory with R-symmetry $K_r$, which has flavor symmetry $K_b\times {K}_l$.
Any choice of an $\Ncal=(0,2)$ subalgebra gives the free theory with 4 chiral and 4 Fermi complex $\Ncal=(0,2)$ superfields.
The flavored elliptic genus in the RR sector of this theory is then
\begin{align} Z_1(\tau | \xi_A, \tilde\zeta_{\tilde A}) &= \Tr_{RR} (-1)^F q^{H_L} \bar{q}^{H_R} \prod_{A=1}^4 a_A^{K_{b,A}}\prod_{\tilde{A}=1}^4 {b}_{\tilde A}^{{K}_{l,\tilde A}}
=  \frac{\theta_1(\tilde\zeta_1)\theta_1(\tilde\zeta_2)\theta_1(\tilde\zeta_3)\theta_1(\tilde\zeta_4)}{\theta_1(\xi_1)\theta_1(\xi_2)\theta_1(\xi_3) \theta_1(\xi_4)} 
\label{eq:08Index}
\end{align} where $\xi_A$ and $ \tilde\zeta_{\tilde A}$ are eigenvalues of flat background gauge fields for $K_b$ and ${K}_l$ corresponding to the Cartan generators $K_{b,A}$ and ${K}_{l,\tilde{A}}$, with 
\be a_A = e^{2\pi i \xi_A}, \quad \tilde b_{\tilde{A}} = e^{2\pi i {\tilde\zeta}_{\tilde{A}}}. \ee We have used the superscript tildes for the $K_l$ Cartan to denote the basis in which the $\mathbf{8}_s$ weights are diagonal. The transformation to the basis in which the $\mathbf{8}_v$ weights are diagonal is given by
\be K_{l,\tilde{A}} =  M_{\tilde A}^{A} K_{l,A},  \quad 
\text{ where } \quad
 M_{\tilde A}^A  =\frac{1}{2} \begin{pmatrix*}[r] 1 & 1&1&~1 \\ 1 & -1&-1&1 \\ -1 & 1&-1&1 \\ -1 & -1&1&1 \end{pmatrix*}. 
\ee

However, this  $K_b\times {K}_l$ flavor symmetry only commutes with the action of a $\Ncal =(0,8)$ superalgebra, and does not respect the full $\Ncal=(8,8)$ supersymmetry of the theory. 
If we insist that $\mathcal{C}$ is indeed an $\Ncal=(8,8)$ supersymmetric theory, there is a single $K=Spin(8)$ R-symmetry, which is not respected by the backgrounds considered above. 
As described in Section \ref{subsec:MSYM elliptic genus setup} and Appendix \ref{reductionofN8superspace}, once an $\Ncal=(0,2)$ subalgebra of the $\Ncal=(8,8)$ algebra is chosen, the supersymmetry generators $\Qcal^{\pm}$ are eigenstates of a corresponding $Spin(2)$ subgroup of $K$, and there is only a $Spin(6) \cong SU(4)$ symmetry commuting with it. 
In this case, we can define an index with fugacities for the $SU(4)$ \lq{}\lq{}flavor\rq{}\rq{} symmetry, which we label $K\rq{}$,
\be Z_1(\tau | \xi\rq{}_B) =  \Tr_{RR} (-1)^F q^{H_L} \bar{q}^{H_R} \prod_{B=1}^3 a\rq{}_B^{K\rq{}_B}, \label{eq:vanishingIndex}
\ee with $K\rq{}_B$ the Cartan generators of $K\rq{}$. But the left-moving fermions and the right-moving supersymmetry generators transform in the same representation, $\mathbf{8}_s$, of $K$. So, for any choice of an $\Ncal=(0,2)$ subalgebra, there will be left-moving fermions which are eigenstates of the $Spin(2)$ R-symmetry, and therefore uncharged under the $SU(4)$ flavor symmetry. The index as defined in \eqref{eq:vanishingIndex} vanishes due to the contributions of these fermion zero modes, as was the case for the free $U(1)$ multiplet as discussed in the paragraph leading up to equation \eqref{eq:U1 index}.

Once again, as is commonly done in the literature, we can remove the contributions from the uncharged fermion zero modes by slightly modifying the index \eqref{eq:vanishingIndex}. This is done by (re)introducing fugacities for symmetries the fermions with problematic fermion zero modes are charged under (so that the modified index has a zero when the fugacities are turned off), taking appropriate derivatives to get rid of the zero, and then turning off the fugacities, as in \cite{Gukov:2004fh} (see also \cite{Cecotti:1992qh}). We can do this by relating \eqref{eq:08Index} to \eqref{eq:vanishingIndex}. First, we identify $K_b$ and $K_l$ diagonally, and write the reduced $\Ncal=(0,8)$ index
\be Z_1(\tau | \xi_A) =  \frac{\theta_1(\frac{\xi_1+\xi_2+\xi_3+\xi_4}{2})\theta_1(\frac{\xi_1-\xi_2-\xi_3+\xi_4}{2})\theta_1(\frac{-\xi_1+\xi_2-\xi_3+\xi_4}{2})\theta_1(\frac{-\xi_1-\xi_2+\xi_3+\xi_4}{2})}{\theta_1(\xi_1)\theta_1(\xi_2)\theta_1(\xi_3) \theta_1(\xi_4)}.  \label{eq:reduced08Index}
\ee 
The $\Ncal=(8,8)$ index \eqref{eq:vanishingIndex} can be computed from \eqref{eq:reduced08Index} by further identifying $K_r$ with $K_b$ and $K_l$ diagonally (so $K_A = K_{b,A} + K_{l,A}+K_{r,A}$), and turning off the fugacity corresponding to the $Spin(2)$ R-symmetry of the $\Ncal=(0,2)$ subalgebra. 
Choosing the $\Ncal=(0,2)$ superalgebra as in Section \ref{subsec:MSYM elliptic genus setup} and equation \eqref{eq:choiceOfN02}, with the R-symmetry generated by $J_R=M_1^A K_A = \frac{1}{2}(K_1+K_2+K_3+K_4)$, we identify
\be K\rq{}_B = M_{B+1}^A K_A,\quad B=1,2,3\ee as the Cartan generators of $K\rq{}$.
Practically, turning off the fugacity for $J_R$ can be realized by having the $\xi_A$ descend to eigenvalues of background flat $SU(4)$-connections, which satisfy the trace constraint \be \xi_1+\xi_2+\xi_3+\xi_4 =0. \ee The $\Ncal=(0,8)$ index \eqref{eq:reduced08Index} has a first-order zero at exactly this constraint due to fermion zero-modes, as it should by our argument above. To remove this zero, we simply take the derivative with respect to $b_1 = \exp(2\pi i \tilde\zeta_1) = \exp(2\pi i \frac{\xi_1+\xi_2+\xi_3+\xi_4}{2})=\sqrt{a_1a_2a_3a_4}$, and set $b_1=1$,
\begin{align}
 \mathcal{I}_1(\tau|\xi_A) &:= \left.-\frac{\partial}{\partial b_1} Z_1(\tau|\xi_A) \right|_{b_1=1}
\\ &= \frac{\eta^3(\tau) \theta_1(\tau| \xi_1+\xi_4)\theta_1(\tau| \xi_2+\xi_4)\theta_1(\tau| \xi_3+\xi_4)}{\theta_1(\tau|\xi_1)\theta_1(\tau|\xi_2)\theta_1(\tau|\xi_3) \theta_1(\tau|\xi_4)}.
\end{align} In this expression, it is understood that the $\xi_A$ satisfy the constraint above. One could explicitly plug in $\xi_4 = -\xi_1-\xi_2-\xi_3$, if desired. We note that this is exactly the index for the (necessarily free) $U(1)$ $\Ncal=(8,8)$ vector multiplet with the vanishing gaugino zero-mode contributions removed, which is a good check that the two definitions of the index for the gauge theory and the sigma model agree.

More generally, for any $\Ncal=(8,8)$ theory, this index is defined as
\begin{align}
 \mathcal{I}(\tau | \xi_A) & =  \left.-\frac{\partial}{\partial b_1} \right|_{b_1=1} \Tr_{RR}(-1)^F q^{H_L} \bar{q}^{H_R} \prod_{A=1}^4 a_A^{K_{A}} 
\\ &= \Tr_{RR} (-1)^F J_R q^{H_L} \bar{q}^{H_R} \prod_{B=1}^3 a\rq{}_B^{K\rq{}_B}.
\end{align}

\paragraph{Fourier expansion of $ \mathcal{I}_1$.}
The index $ \mathcal{I}_1$ enjoys a number of very special properties. For definiteness, we will solve for the $SU(4)$ (or, really, $SL(4,\CC)$) constraint by setting $\xi_4= -\xi_1-\xi_2-\xi_3$ explicitly in this section.
\begin{itemize}
	\item (Abelian function.) $\mathcal{I}_1$ is holomorphic in $\tau \in \mathbb{H}/SL(2,\Z)$ (including at the cusp $q=0$ or $\tau=i\infty$), and meromorphic in each $\xi_A\in \CC/(\Z+\tau \Z)$. Moreover, $\mathcal{I}_1$ is doubly periodic in each $\xi_A$ under translations by the lattice $\Z+\tau \Z$, i.e. 
\begin{equation}\mathcal{I}_1(\tau|\xi + \Omega \cdot n) = \mathcal{I}_1(\tau|\xi), \end{equation}
where $n\in \Z^6$ and $\Omega$ is the period matrix \begin{equation} \Omega = \begin{pmatrix} 1 & \tau & 0 &0&0&0 \\  0&0&1 & \tau & 0 &0 \\ 0 &0&0&0 &1 & \tau  \end{pmatrix}. \label{eq:period matrix}\end{equation} 
	\item (Symmetric function.) $\mathcal{I}_1$ is symmetric in $\xi_A$.
	\item (Modularity.) $\mathcal{I}_1$ is modular invariant, i.e. under $SL(2,\Z)$ transformations $\tau \rightarrow \frac{a\tau + b}{c\tau +d}$, we have, \begin{align} \mathcal{I}_1 \left. \left(\frac{a\tau+b}{c\tau+d} \right| \frac{\xi_A}{c\tau+d}\right) = \mathcal{I}_1(\tau | \xi_A), \qquad \begin{pmatrix} a & b \\ c & d \end{pmatrix} \in SL(2,\Z).\end{align} 
\end{itemize}

It follows from these properties that $\mathcal{I}_1$ is an honest map $(\mathbb{H}/ SL(2,\Z) )\times (\CC/\Z+\tau\Z)^3 \rightarrow \CC$, and also a $3$ variable Jacobi form (function) of weight $0$ and index $(0,0,0)$. The periodicity in $\tau \rightarrow \tau+1$ and $\xi_A\rightarrow \xi_A+1$ allows for a Fourier expansion, of the form
 \be \mathcal{I}_1(\tau | \xi_A) =   \sum_m q^m f_m(\xi) = \sum_{m\ge0, l} c(m,l) q^m \prod_A a_A^{l_A}. \ee Since the function is holomorphic in $q$, the coefficients $f_m(\xi)$ of $q^m$ are unique and well-defined. But since the $f_m$ are meromorphic functions themselves, they might have multiple Fourier expansions. For example, we can easily determine \be \begin{split} \left.\mathcal{I}_1\right|_{q=0} (\xi) &:=\mathcal{I}_1(\tau=i\infty | \xi) = \frac{(1-a_1a_2)(1-a_1a_3)(1-a_2a_3)}{(1-a_1)(1-a_2)(1-a_3)(1-a_1a_2a_3)} 
 \\ &= 1+\frac{a_1}{1-a_1}+\frac{a_2}{1-a_2}+\frac{a_3}{1-a_3} - \frac{a_1a_2a_3}{1-a_1a_2a_3}. \end{split} \ee The function $\left.\mathcal{I}_1\right|_{q=0} (\xi)$ has different Fourier expansions in different regions of convergence of the $a_A$.
Now, we can use the periodicity in $\xi_A \rightarrow \xi_A+\tau$ to find a recursion relation for $c(m,l_A)$, which, when combined with modular invariance, determines $\mathcal{I}_1$ completely given $\left.\mathcal{I}_1\right|_{q=0} :=\mathcal{I}_1(\tau=i\infty | \xi) $.  Explicitly, we have
\be \mathcal{I}_1(\tau | \xi) = \left.\mathcal{I}_1\right|_{q=0} (\xi)
+ \sum_{m=1}^\infty q^m \sum_{s|m} \chi(s\xi) \label{eq:R8IndexFourierExpansion}\ee where $\chi(\xi)$ is the $SL(4,\CC)$ character
\be \chi(\xi_A) =  \chi_{\Box}(\xi_A)-  \chi_{\bigwedge^3{\Box}}(\xi_A) = a_1 + a_2 + a_3 +a_4 - \frac{1}{a_1} - \frac{1}{a_2} - \frac{1}{a_3} - \frac{1}{a_4}. \ee 

To see this, 
note that the periodicity of $\mathcal{I}_1$ under $\xi_1 \rightarrow \xi_1 + \tau$ implies the identity, 
\be  \sum_{m\ge0, l_A} c(m,l_1,l_2,l_3) q^m a_1^{l_1}a_2^{l_2}a_3^{l_3} =  \sum_{m\ge 0, l_A} c(m,l_A) q^{m+l_1} a_1^{l_1}a_2^{l_2}a_3^{l_3}, \ee
and similarly for $\xi_2$ and $\xi_3$. To retain a holomorphic series expansion in $q$, we must choose $c(0,l)$ to be the coefficients of the expansion of $\left.\mathcal{I}_1\right|_{q=0}$ in positive powers of $a_A$, i.e. the expansion convergent in the region $|a_A|<1$. From here, for each $A=1,2,3$ and $m\ge0$, we infer the following relations
\begin{equation} c(m,l_1,l_2,l_3) = \begin{cases} 0 & \mbox{if }m>0 \mbox{ and, } m+l_A<0 \mbox{ or } m-l_A<0, \\ c(m+l_A,l_1,l_2,l_3) & \mbox{if } m+l_A > 0.\end{cases} \end{equation}
The case with $l_A<0$ such that $m+l_A = 0$ should be handled with more care. In that case, for say $A=1$, we have
\begin{equation}  \left.\mathcal{I}_1\right|_{q=0}(\xi) = \sum_{l_1 \le 0,l_2,l_3} c(-l_1,l_1,l_2,l_3) a_1^{l_1}a_2^{l_2}a_3^{l_3}, \end{equation} which determines  $c(-l_A,l_1,l_2,l_3) = \tilde c(0,l_1,l_2,l_3)$ where $\tilde c$ are the coefficients of $\left.\mathcal{I}_1\right|_{q=0}$ in the expansion with negative powers of $a_A$. Putting it together, we have\footnote{We should note that for the general case, the first two cases should be generalized to hold for the conditional $n^Al_A =m$ for some integers $n^A$, rather than just $l_A|m$. But for the specific case of $\mathcal{I}_1$, since $c(0,l)$ is only nonzero when $l=(l_1,l_2,l_3)$ is of the form $(l,0,0)$, $(0,l,0)$, $(0,0,l)$, or $(l,l,l)$, the notions coincide.}
\begin{equation} c(m,l_1,l_2,l_3) = \begin{cases} c(0,l_1,l_2,l_3) & \mbox{if } l_A> 0 \mbox{ and } l_A|m \mbox{ for some }A, \\ \tilde{c}(0,l_1,l_2,l_3) & \mbox{if } l_A<0 \mbox{ and } l_A|m\mbox{ for some }A, \\ c(m,0,0,0) & \mbox{if } l_A=0 \mbox{ for all }A,\\0 & \mbox{otherwise}.\end{cases} \end{equation} 
The only coefficients that are not determined by these relations are those of the form $c(m,0,0,0)$, implying that the function is determined up to a holomorphic function of $q$. Requiring the function to be modular invariant fixes this ambiguity, since the only holomorphic modular invariant functions are constants. For $\mathcal{I}_1$, $c(m,0,0,0) = 0$ for $m>0$, and we obtain \eqref{eq:R8IndexFourierExpansion}.

It is important to note that our discussion above proves that if any Abelian, modular invariant function $f(\tau|\xi)$ with the same period matrix $\Omega$ as $\mathcal{I}_1(\tau|\xi)$ agrees with $\mathcal{I}_1$ at $q=0$, then it must equal $\mathcal{I}_1$. More generally, we have the following result.




\begin{lemma} \label{lemma:ModularInvariantAbelianFunction} Let $f(\tau | \xi_A)$ be a modular invariant, Abelian function with periods $1$ and $\tau$ for each $\xi$, holomorphic in $\tau$ (including at the cusp, $q=0$) and meromorphic in $\xi_A$. Then $f(\tau | \xi_A)$ is completely determined by $\left.f\right|_{q=0}(\xi_A) = f(\tau=i\infty | \xi_A)$. 
\end{lemma}

A particularly useful class of such functions for us turn out to be $\mathcal{I}_1(\tau|N \xi)$, which satisfy the same properties as $\mathcal{I}_1(\tau|\xi)$.

\subsection{Elliptic genus of the $\Sym^N(\R^8)$ sigma model}
There are various equivalent methods of computing the partition function $Z_N$ of a symmetric product theory given the partition function of the base theory $Z_1$. We list three prominent methods here.
\begin{itemize}
	\item Summing over $\mathbb{S}_N$ connections and twisted sectors
	\be
	Z_N = \frac{1}{|\mathbb{S}_N|} \sum_{gh=hg} (Z_1^N)^{g,h}
	\ee
	\item The DMVV formula \cite{Dijkgraaf:1996xw}
	\be \mathcal{Z} :=  1+ \sum_{N\ge 1} p^N Z_N(q,\vec a) = \prod_{n> 0, m\ge 0,\vec l} \frac{1}{(1-p^n q^m \vec{a}^{\vec l})^{c(nm,\vec l)}}. \ee
	\item Hecke operators \cite{Gukov:2004fh}
	\be \log \mathcal{Z} = \sum_{M=1}^\infty p^M T_M Z_1, \ee so in particular
	\be Z_N = T_N Z_1 + \dots +\frac{1}{N!} (T_1 Z_1)^N
	\ee where \be T_M Z_1(\tau | \vec \xi) := \frac{1}{M} \sum_{d|M,d=1}^M \sum_{b=0}^{M/d-1} Z_1 \left( \left. \frac{d \tau +b}{M/d} \right| {d}\vec \xi\right).\ee
\end{itemize}
The Hecke operators turn out to be the most straightforward to extract a closed-form expression for $Z_N$, given one for $Z_1$. For the index we are interested in, we need to perform the \lq\lq{}index operation\rq\rq{} to remove zero-mode contributions,
\be \mathcal{I}_N := \left.-\frac{\partial}{\partial b_1} \right|_{b_1=1} Z_N, \ee like we did to obtain $\mathcal{I}_1$. Analogous to the case in \cite{Gukov:2004fh}, only the term linear in $Z_1$ survives this operation, as all the other terms have zeroes of order greater than $1$ at $b_1 = 1$. Thus,
\begin{align}  \mathcal{I}_N &=  \left.-\frac{\partial}{\partial b_1} \right|_{b_1=1} T_N Z_1 = \frac{1}{N} \sum_{d|N} \sum_{b=0}^{N/d-1} d \;  \mathcal{I}_1 \left( \left. \frac{d \tau +b}{N/d} \right| {d}\xi_A \right). \label{eq:indexSymNHecke} \end{align}
Specializing to the sigma model into $\Sym^N(\R^8)$, turns out we can simplify further, \be  \mathcal{I}_N= \sum_{d|N} \mathcal{I}_1(\tau | d \xi_A). \label{eq:indexSymN} \ee
This last simplification is nontrivial, but can be seen in two ways. One can notice that the $q=0$ piece of the two expressions in \eqref{eq:indexSymNHecke} and \eqref{eq:indexSymN} agree, and they are both periodic functions on $(\CC/\Z+\tau\Z)^3$; therefore they are equal by Lemma \ref{lemma:ModularInvariantAbelianFunction}. Alternatively, one can directly compute from the Fourier expansion:
\be \begin{split}
 \frac{1}{N} \sum_{d|N} \sum_{b=0}^{N/d-1} d \;  \mathcal{I}_1 \left( \left. \frac{d \tau +b}{N/d} \right| {d}\xi_A \right) &=  \sum_{d|N} \left.\mathcal{I}_1\right|_{q=0}\left(d \xi_A \right) + \sum_{d|N} \sum_{m=1}^\infty q^{d m} \sum_{s| \frac{N}{d} m} \chi(sd\xi_A)
\\ &=  \sum_{d|N} \mathcal{I}_1\left(\tau=i\infty | d \xi_A \right) + \sum_{k=1}^\infty q^{k}  \sum_{d\rq{}|N} \sum_{s\rq{}| k} \chi(s\rq{}d\rq{}\xi_A)
\\ &= \sum_{d|N} \mathcal{I}_1(\tau | d \xi_A). 
\end{split} \ee


\section{Conclusions and future directions}

We have computed the elliptic genera of the $SU(N)/\Z_K$ $\msym_2$ and $U(N)$ $\msym_2$ with and without the $B$ field, with each corresponding choice of the discrete $\theta$ angle, and matched it with the elliptic genus of a corresponding $\Ncal=(8,8)$ sigma model into a symmetric orbifold of $\R^8$, which we claim describes the IR fixed point in that sector. While the main focus of this work was in answering questions about the vacua of $\msym_2$, the elliptic genera we have computed as part of our analysis are interesting objects in their own rights. For example, they are related to the supersymmetric partition function of the free second quantized Type IIA string as explored in \cite{Dijkgraaf:1996xw}, if one performs the sum over the string winding number $N$;
\begin{align}
\mathcal{Z}_0 (\tau,\sigma|\xi) =1+ \sum_{N\ge 1} p^N \mathcal{I}_{N}(\tau|\xi),
\end{align} where $p=e^{2\pi i \sigma}$. One needs to modify this expression with an appropriate factor to obtain the T-duality invariant partition function $\mathcal{Z}(\tau,\sigma|\xi)$ \cite{Dijkgraaf:1996xw}. T-duality exchanges sting winding number and oscillator number, so acts by interchanging $p$ and $q$, which can be used to determine $\mathcal{Z}$. One could try to extract information about the strongly coupled limit of the string, which is M-theory, using the topological invariance of this function. It would also be an interesting question to understand the automorphic properties of $\mathcal{Z}$, \emph{a la} \cite{Borcherds:1995aa}. One might also consider replacing $\mathcal{I}_N$ with the full D1-brane index $\mathcal{I}_{U(N)+B}$, which in the Type IIA picture sums over the bound states with D0-branes as well.

This work was inspired by the 4d-2d correspondence explored in \cite{Gadde:2013sca}, as well as by recent developments in the computation of flavored elliptic genera for 2d gauge theories. In particular, $\msym_2$ can be obtained by considering M5-branes on a four-dimensional torus $T^4$ and letting the volume of the $T^4$ shrink to zero. On the other hand, considering M5-branes on $T^6 = T^2 \times T^4$, and compactifying first on the $T^2$ factor taken to be the worldvolume of the $\msym_2$, one obtains 4d $\Ncal=4$ SYM. Following the general idea of \cite{Gadde:2013sca}, the elliptic genus of $\msym_2$ is then related to the Vafa-Witten partition function of the 4d $\Ncal=4$ theory on $T^4$, as well as to an appropriate supersymmetric partition function of the 6d $\Ncal=(2,0)$ theory on $T^6$. We will be exploring this relation in upcoming work.


\acknowledgments{
The author would like to thank Sergei Gukov for suggesting this problem, and for his guidance. The author would also like to thank Abhijit Gadde and Du Pei for valuable discussions, and especially Ingmar Saberi for discussions and comments on a draft of this work. This material is based upon work supported by the U.S. Department of Energy, Office of Science, Office of High Energy Physics, under Award Number DE-SC0011632.
}



\appendix

\section{Action and supersymmetry transformations of $\msym_2$}\label{apx:action and susy}

\subsection{Dimensional reduction from 10d to 2d} \label{apx:dim red and susy}

The Lagrangian for the $\Ncal = (8,8)$ super Yang-Mills theory in 2 dimensions can be obtained by dimensionally reducing the 10 dimensional $\Ncal=1$ SYM action
\be
\int d^{10} x \Tr \left( -\frac{1}{4} F_{MN} F^{MN} +\frac{i}{2} \bar{\Theta} \Gamma^M D_M \Theta\right)
\ee where 
\ba D_M &= \partial_M + i g [ A_M,\cdot]
\\ F_{MN} &= \frac{1}{ig} [D_M,D_N] = \partial_M A_N - \partial_N A_M +ig[A_M,A_N].
\ea
The dimensionally reduced Lagrangian is~\cite{F.Antonuccio:1998aa,Dijkgraaf:1997vv}
\be
\Lcal = \Tr\left( -\frac{1}{2} (D_\mu X^i)^2 + i\chi^T \slashed{D} \chi -\frac{1}{4} F_{\mu\nu}^2 +\frac{g^2}{4}[X^i,X^j]^2 -\sqrt{2} g \chi_L^T \gamma_i[X^i,\chi_R] \right). \label{eq:msym lag apx}
\ee 

We will summarize the derivation presented in~\cite{F.Antonuccio:1998aa}, but adopt a ``mostly plus'' metric signature in contrast. We use the 10 dimensional metric \be g_{MN} = \eta_{\mu \nu} \oplus \delta_{ij} \ee where $\mu,\nu=0,9$, and $i,j=1,2,\dots,8$, and $\eta_{\mu\nu}={\rm diag}(-1,+1)$. We can write the following 10d Majorana-basis (purely imaginary) gamma matrices satisfying $\{ \Gamma^M,\Gamma^N\} = -2 g^{MN}$ 
\be \begin{split} \Gamma^0 &= \sigma_2 \otimes I_{16}
\\ \Gamma^i &= i \sigma_1 \otimes \gamma^i
\\ \Gamma^9 &=i \sigma_1 \otimes \gamma^9, \end{split} 
\qquad 
\begin{split}
\gamma^i &= \mattwo{0}{\beta_i}{\beta_i^T}{0}, 
\\ \gamma^9 &=\mattwo{I_8}{0}{0}{-I_8},
\end{split}
\ee where the $\sigma_a$ are the usual Pauli matrices,
\be \sigma_1=\mattwo{0}{1}{1}{0}, \quad \sigma_2 = \mattwo{0}{-i}{i}{0}, \quad \sigma_3 = \mattwo{1}{0}{0}{-1}, \ee and the $\gamma^i$ are $16\times 16$ $SO(8)$ gamma matrices of the reducible $\mathbf{8}_s\oplus\mathbf{8}_c$ representation, with the $\beta_i$ satisfying $\{\beta_i,\beta^T_j\}=2\delta_{ij}$. The 10d spinor $\Theta$ is Majorana, and has real components in the Majorana basis we have chosen above, thus we can identify the charge conjugation matrix $\mathcal C = - \Gamma^0$. $\Theta$ also satisfies the Weyl condition $\Theta = \Gamma^{11} \Theta$, where $\Gamma^{11} = \Gamma^0 \cdots \Gamma^9 = \sigma_3 \otimes I_{16}$ is the  10d chirality matrix, which allows us to write $\Theta = (\chi,0)^T$. The 8d chirality matrix $\gamma^9$ allows us to decompose further as $\chi = (\chi_L,\chi_R)$. 

Dimensionally reducing on the $1,2,\dots,8$ directions, we define scalars $ X^i:= A^i$, 
and obtain the action
\be \begin{split}
S_{\msym_2} &= 
\int dx^2 \Tr\left( -\frac{1}{2} (D_\mu X^i)^2 + \frac{i}{2} \chi_L^T (D_0+D_9) \chi_L + \frac{i}{2} \chi_R^T (D_0-D_9) \chi_R-\frac{1}{4} F_{\mu\nu}^2 \right.\\ & \qquad \left. +\frac{g^2}{4}[X^i,X^j]^2 - g\chi_L^\alpha \gamma^i_{\alpha\dot \beta}[X^i,\chi_R^{\dot \beta}] \right). \label{eq:msymaction}
\end{split}
\ee

We are interested in the theory with gauge group $U(N)$ or $SU(N)$. The scalars $X^i$ and the fermions $\chi = (\chi_L^\alpha,\chi_R^{\dot{\alpha}})$ are in the adjoint of the gauge group. The Lagrangian manifestly possesses a $Spin(8)$ R-symmetry, interpreted as rotations in the $8$ transverse directions, under which the scalars $X^i$ and the spinors $\chi_L^\alpha$, and $\chi_R^{\dot{\alpha}}$ transform in the $\mathbf{8}_v$, $\mathbf{8}_s$, and $\mathbf{8}_c$ representations, respectively. 

The supersymmetry transformations can be deduced from the 10d SYM transformations \cite{Brink:1976bc}:
\ba \delta A_M &= i \bar \varepsilon \Gamma_M \Theta 
\\ \delta \Theta &= \Gamma_{MN} F^{MN} \varepsilon. \label{eq:10d susy}
\ea
After dimensional reduction, they are given by
\ba \delta A_\mu &= i\varepsilon^T \Gamma^0 \Gamma_\mu \chi
\\ \delta X^i &= i \varepsilon_L^{\alpha} \gamma^i_{\alpha\dot{\alpha}} \chi_R^{\dot \alpha} + i \varepsilon_R^{\dot\alpha} \gamma^i_{\dot \alpha \alpha} \chi_L^{\alpha}
\\ \delta \chi_L^\alpha &= 4c\left[(+F_{09} \delta_{\alpha \beta} -\frac{ig}{2}  [X_i,X_j] \gamma^i_{\alpha \dot\rho}\gamma^j_{\dot\rho \beta} ) \varepsilon^\beta_L + (D_0 +D_9) X_i \gamma^i_{\alpha \dot \beta} \varepsilon^{\dot \beta}_R\right] 
\\ \delta \chi_R^{\dot \alpha} &= 4c\left[(- F_{09}\delta_{\dot\alpha \dot\beta} -\frac{ig}{2} [X_i,X_j]\gamma^i_{\dot \alpha \rho}\gamma^j_{\rho \dot\beta}) \varepsilon^{\dot\beta}_R +  (D_0 - D_9) X_i \gamma^i_{\dot \alpha \beta}\varepsilon^{\beta}_L\right] 
\ea
where $c$ is the constant in $\Gamma^{MN} = c [\Gamma^M,\Gamma^N]$, and is determined as $c=\frac{1}{4i}$ by imposing $\Gamma^{MN} \Gamma_{MN} = \binom{10}{2}$. In the $U(N)+B$ theory, one should replace $F_{09}$ with the generalized field strength $\mathcal{F}_{09}$.

%

\subsection{Supersymmetry subalgebras and superspace formulation} \label{reductionofN8superspace}

For the purpose of computing the index of $\msym_2$, it is convenient to express fields and the Lagrangian in $\Ncal =(0,2)$ or $\Ncal=(2,2)$ superspace. This can be done by considering the representations of the fields and supersymmetries under the $Spin(8)$ R-symmetry. The 16 supersymmetry generators $(\varepsilon_L^{\dot \alpha},\varepsilon_R^\alpha)$ are in the representation $\mathbf{8}_c \oplus \mathbf{8}_s$ of $Spin(8)$. A choice of a $\Ncal =(0,2)$ subalgebra of the supersymmetry algebra is generated by $\varepsilon_R^\pm:= \varepsilon_R^1 \pm i \varepsilon_R^2$ corresponding to a pair of antiparallel weights of the $\mathbf{8}_s$ representation. Letting $\{\pm e_i\}\subset \mathfrak{h}^*$ be the weights of the fundamental representation $\mathbf{8}_v$, we pick the two weights $\pm r$ of $\mathbf{8}_s$ where  
\be r:=\foh (e_1+e_2+e_3+e_4). \label{eq:choiceOfN02}\ee Note that $\pm r$ are eigenvalues for the action of the Cartan generator $J=\foh(K_1 + K_2 + K_3 +K_4)$ on the weightspaces of $\pm r$, where $e_i(K_k)=\delta_{ik}$.
With this choice, the $Spin(8)$ representations reduce as 
\be \begin{split} \mathbf{8}_s &\rightarrow \mathbf{1}_{+1} \oplus \mathbf{6}_0 \oplus \mathbf{1}_{-1}
\\ \mathbf{8}_c &\rightarrow \mathbf{4}_{-\foh} \oplus \bar{\mathbf{4}}_{+\foh}
\\ \mathbf{8}_v &\rightarrow \mathbf{4}_{+\foh} \oplus \bar{\mathbf{4}}_{-\foh}
\end{split}
\label{eq:SO8toSO2timesSU4branching}
\ee
under the decomposition $U(1)_R \times SU(4) \cong Spin(2) \times Spin(6) \subset Spin(8)$, where $U(1)_R$ is generated by $J$. The supersymmetry generators are now $(\varepsilon_L^{\dot\alpha}, \varepsilon_R^\alpha) = (\varepsilon_L^A,(\varepsilon_L)_A, \varepsilon_R^\pm,\varepsilon_R^{AB})$, where $A,B = 1,2,3,4$ are $SU(4)$ indices for the fundamental representation $\mathbf{4}$. The field content of the theory is organized into $\Ncal=(0,2)$ superfields as in \eqref{eq:02 superfields} in the main text, with the Lagrangian given by the standard D-terms and the superpotential \eqref{eq:Jterm superpotential}.

To get an $\Ncal = (2,2)$ subalgebra, one can to pick $l:= \foh(e_1+e_2+e_3-e_4)$. Then, the vector and axial R-symmetries are determined by 
\be
R_V = r+l = e_1+e_2+e_3, \qquad R_A = r-l = e_4
\ee
This choice further decomposes the R-symmetry to $ U(1)_R \times U(1)_L \times SU(3) \subset Spin(8)$, with the representations decomposing as
\be \begin{split} \mathbf{8}_s &\rightarrow \mathbf{1}_{+1,+\foh} \oplus \mathbf{3}_{0,+\foh} \oplus \bar{\mathbf{3}}_{0,-\foh} \oplus \mathbf{1}_{-1,-\foh}
\\ \mathbf{8}_c &\rightarrow \mathbf{3}_{-\foh,0} \oplus \mathbf{1}_{-\foh,-1} \oplus \bar{\mathbf{3}}_{+\foh,0} \oplus \mathbf{1}_{+\foh,+1}
\\ \mathbf{8}_v &\rightarrow \mathbf{3}_{+\foh,+\foh} \oplus \mathbf{1}_{+\foh,-\foh} \oplus \bar{\mathbf{3}}_{-\foh,-\foh} \oplus \mathbf{1}_{-\foh,+\foh}
\end{split}
\ee
The supersymmetries are generated by $(\varepsilon_L^\pm,\varepsilon_L^A,(\varepsilon_L)_A, \varepsilon_R^\pm,\varepsilon_R^A,(\varepsilon_R)_A)$. In $\Ncal=(2,2)$ superspace, the $SU(3)$ singlets correspond to the components of the vector multiplet $\tilde \Sigma$ and its conjugate, and the $\mathbf{3} \oplus \bar{\mathbf{3}}$ correspond to the compontents of the chiral fields $\tilde \Phi^B$ and its conjugate, with the superpotential as in \eqref{eq:Fterm superpotential}.

\bibliographystyle{JHEP}
\bibliography{M5partition}


\end{document}